\documentclass[prb,twocolumn,amsmath,amssymb]{revtex4}

\usepackage{graphicx}% Include figure files
\usepackage{bm}

\begin{document}
\title{Principle of Maximum Entropy Applied to
Rayleigh-B\'enard Convection}

\author{Takafumi Kita}
%\homepage{http://phys.sci.hokudai.ac.jp/~kita/index-e.html}
\affiliation{Department of Physics, Hokkaido University, Sapporo 060-0810, Japan}

\date{\today}

\begin{abstract}
A statistical-mechanical investigation is performed on 
Rayleigh-B\'enard convection of a dilute classical gas
starting from the Boltzmann equation.
We first present a microscopic derivation of basic hydrodynamic equations
and an expression of entropy appropriate for the convection.
This includes an alternative justification 
for the Oberbeck-Boussinesq approximation.
We then calculate entropy change through the convective transition
choosing mechanical quantities as
independent variables.
Above the critical Rayleigh number,
the system is found to evolve from the heat-conducting uniform state towards the 
convective roll state with monotonic increase of entropy on the average.
Thus, the principle of maximum entropy proposed 
for nonequilibrium steady states in a preceding paper
[T.\ Kita:\ J.\ Phys.\ Soc.\ Jpn.\  {\bf 75} (2006) 114005]
is indeed obeyed in this prototype example.
The principle also provides a natural explanation for 
the enhancement of the Nusselt number in convection.
\end{abstract}

\keywords{% 
Rayleigh-B\'enard convection, Nusselt number, Boltzmann equation, Entropy,
Oberbeck-Boussinesq approximation}

\maketitle

\section{\label{sec:intro}Introduction}

In a preceding paper,\cite{Kita06} we have proposed a principle of maximum entropy
for nonequilibrium steady states:
{\em The state which is realized most probably 
among possible steady states without time evolution
is the one that makes entropy maximum as a function
of mechanical variables.}
We here apply it to Rayleigh-B\'enard 
convection of a dilute classical
gas to confirm its validity.

Rayleigh-B\'enard convection is a prototype of nonequilibrium steady 
states with pattern formation, 
and extensive studies have been carried out 
to clarify it.\cite{Chandrasekhar61,Busse78,Behringer85,Croquette89,
CH93,Koschmieder93,deBruyn96,BPA00}
However, no calculation seems to have been performed 
on entropy change through the nonequilibrium ``phase transition,''
despite the fact that entropy is the key concept in equilibrium thermodynamics and
statistical mechanics.
There may be at least four reasons for it.
First, there seems to have been no established expression of nonequilibrium
entropy.
Second, the standard starting point to describe Rayleigh-B\'enard convection 
is a set of deterministic equations for the particle, momentum and energy flows,
with all the thermodynamic effects pushed 
into phenomenological parameters of the equations.\cite{Chandrasekhar61,CH93}
Third, one additionally adopts the Oberbeck-Boussinesq 
approximation
to the equations in the conventional treatment.\cite{Chandrasekhar61,CH93,Joseph71}
Despite many theoretical efforts 
over a long period,\cite{SV60,Mihaljan62,CV75,GG76,RRS96} 
a well-accepted systematic justification 
for it seems still absent, thereby
preventing a quantitative estimation of entropy change.
Fourth, there is ambiguity on what to choose as independent variables of 
entropy for open systems.

In a preceding paper,\cite{Kita06} 
we have derived an expression of nonequilibrium entropy
together with the evolution equations
for interacting bosons/fermions.
We here apply them to a classical gas of the dilute high-temperature limit
where the evolution equations reduce to the Boltzmann equation.
We carry out a microscopic derivation of the hydrodynamic equations 
for the particle, momentum and energy densities
(i.e., the basic conservation laws)
from the Boltzmann equation.
We then provide a systematic justification for the Oberbeck-Boussinesq approximation
to describe the convection.
With these preliminaries, we perform a statistical-mechanical calculation 
of entropy change through the convective transition
by choosing mechanical quantities as independent variables.
It is worth pointing out that classical gases have been used extensively
for detailed experiments on Rayleigh-B\'enard convection over the last
two decades.\cite{Croquette89,deBruyn96,BPA00} 
Thus, quantitative comparisons between 
theory and experiment are possible here.

This paper is organized as follows.
Section \ref{sec:EqMot} derives (i)
equations of motion for the particle, momentum and energy densities
and (ii) an explicit expression for the distribution function $f$,
both starting from the Boltzmann equation.
Section \ref{sec:RB} reduces the equations of \S\ref{sec:EqMot}
in a way appropriate to treat Rayleigh-B\'enard convection of a dilute classical gas.
This includes a systematic justification of
the Oberbeck-Boussinesq approximation and a derivation of the expression
of entropy for convection.
Section \ref{sec:Fourier} transforms the equations of 
\S\ref{sec:RB} into those suitable for periodic structures with
the stress-free boundaries.
Section \ref{sec:num} presents numerical results obtained by solving the equations
of \S\ref{sec:Fourier}. 
It is shown explicitly that the principle of maximum entropy
is indeed obeyed by the convection.
Concluding remarks are given in \S\ref{sec:summary}.

\section{\label{sec:EqMot}
Distribution function, conservation laws and entropy}

\subsection{The Boltzmann equation and entropy}

We shall consider a monatomic dilute classical gas under gravity.
This system may be described by the Boltzmann equation:\cite{CC90,HCB54}
\begin{equation}
\frac{\partial f}{\partial t}+
\frac{{\bm p}}{m}\cdot\frac{\partial f}{\partial{\bm r}}
-mg\frac{\partial f}{\partial p_{z}}={\cal C} \, .
\label{Boltzmann}
\end{equation}
Here $f\!=\!f({\bm p},{\bm r},t)$ is the distribution function,
$t$ the time, ${\bm p}$ the momentum, 
$m$ the mass, ${\bm r}$ the space coordinate, 
and $g$ the acceleration of gravity.
With a unified description of classical and quantum statistical mechanics in mind,
we choose the normalization of $f$
such that it is dimensionless and 
varies between $0$ and $1$ for fermions.
The collision integral ${\cal C}$ is given explicitly by
\begin{eqnarray}
&&\hspace{-5mm}
{\cal C}({\bm p},{\bm r},t)
\nonumber \\
&&\hspace{-5mm}=\hbar^{2}\!\int\!\!\frac{{\rm d}^{3}p_{1}}{(2\pi\hbar)^{3}}
\!\int\!\!\frac{{\rm d}^{3}p'}{(2\pi\hbar)^{3}}
\!\int\!\!\frac{{\rm d}^{3}p_{1}'}{(2\pi\hbar)^{3}}
|V_{{\bm p}'-{\bm p}}|^{2}(f_{{\bm p}'}f_{{\bm p}_{1}'}\!-\!
f_{{\bm p}}f_{{\bm p}_{1}})
\nonumber \\
&&\hspace{-1.5mm}
\times(2\pi)^{4}
\delta(E_{{\bm p}'}\!+\!
E_{{\bm p}_{1}'}\!-\!E_{{\bm p}}\!-\!E_{{\bm p}_{1}})
\delta({\bm p}'+
{\bm p}_{1}'\!-\!{\bm p}\!-{\bm p}_{1}) \, ,
\label{C1}
\end{eqnarray}
where $V_{\bm q}$ is Fourier transform of the interaction potential
and $E_{{\bm p}}\!=\!p^{2}/2m$.

Equation (\ref{Boltzmann}) 
also results from eq.\ (63) of ref.\ \onlinecite{Kita06}
as follows:
(i) approximate the spectral function as
$A({\bm p}\varepsilon,{\bm r}t)\!=\!2\pi \delta(\varepsilon\!-\!E_{{\bm p}}\!-\!mgz)$;
(ii) substitute the second-order self-energy of eq.\ (66)
into the collision integral of eq.\ (64); 
(iii) take the high-temperature limit;
and (iv) integrate eq.\ (63) over $\varepsilon$ to obtain an equation for
\begin{equation}
f({\bm p},{\bm r},t)\equiv\!\int_{-\infty}^{\infty}\!\frac{{\rm d}\varepsilon}{2\pi}
A({\bm p}\varepsilon,{\bm r}t)\phi({\bm p}\varepsilon,{\bm r}t)\, .
\label{f-def}
\end{equation}
This whole procedure amounts to treating the interaction potential
only as the source of dissipation in the dilute high-temperature limit, 
thus neglecting completely its influence on the density of states, i.e.,
on the real part of the self-energy.

With a change of variables ${\bm p}_{1}\!=\!{\bm p}\!+\!{\bm q}$, 
${\bm p}'\!=\!{\bm P}\!-\!{\bm q}'/2$ and ${\bm p}_{1}'\!=\!{\bm P}\!+\!{\bm q}'/2$
in eq.\ (\ref{C1}), the collision integral is transformed into
\begin{equation}
{\cal C}=\!\int\!\!\frac{{\rm d}^{3}q}{(2\pi\hbar)^{3}}\,\frac{q}{m}\!\int\!
{\rm d}\sigma\, [f_{{\bm p}+({\bm q}-{\bm q}')/2}
f_{{\bm p}+({\bm q}+{\bm q}')/2}\!-\!f_{{\bm p}}f_{{\bm p}+{\bm q}}] \, ,
\label{C2}
\end{equation}
where ${\rm d}\sigma\!=\! {\rm d}\Omega_{{\bm q}'}\int\!{\rm d}q'
\delta(q'\!-\!q)\left[m|V_{({\bm q}'-{\bm q})/2}|/4\pi\hbar^{2}\right]^{2}$
is the differential cross section in the center-of-mass
coordinate of the scattering with ${\rm d}\Omega_{{\bm q}'}$ denoting
the infinitesimal solid angle.
We shall use the contact interaction
with no ${\bm q}$ dependence in $V_{{\bm q}}$,
where ${\rm d}\sigma$ reduces to
\begin{equation}
{\rm d}\sigma=a^{2}{\rm d}\Omega_{{\bm q}'}\int\!{\rm d}q'
\delta(q'\!-\!q) \, ,
\label{dsigma}
\end{equation}
with $a\!\equiv\! m|V|/4\pi\hbar^{2}$ denoting the scattering length.
Now, the differential cross section acquires the form 
of the two-particle collision between hard-sphere particles
with radius $a$.\cite{LL-M}

Entropy per unit volume is given in terms of $f$ by
\begin{equation}
S=-\frac{k_{\rm B}}{\cal V}\int\frac{{\rm d}^{3}r{\rm d}^{3}p}{(2\pi\hbar)^{3}}\, 
f(\log f-1) \, ,
\label{S}
\end{equation}
with ${\cal V}$ denoting the volume.
This expression also results from eq.\ (69) of ref.\ \onlinecite{Kita06}
for entropy density
by: (i) adopting the quasiparticle approximation 
$A({\bm p}\varepsilon,{\bm r}t)\!=\!2\pi \delta(\varepsilon\!-\!E_{{\bm p}}\!-\!mgz)$;
(ii) defining $f$ by eq.\ (\ref{f-def}) above;
(iii) taking the high-temperature limit;
and (iv) performing integration over ${\bm r}$.
The term $-1$ in the integrand of eq.\ (\ref{S}) is absent
in the Boltzmann $H$-function\cite{CC90} 
but naturally results from the procedure (iii) above.
Indeed, eq.\ (\ref{S}) reproduces the correct expression of 
entropy in equilibrium.\cite{Reif}

\subsection{\label{subsec:conserve}Conservation laws}

We next consider conservation laws which originate from the Boltzmann equation.
The basic physical quantities relevant to them are
the density $n({\bm r},t)$, the velocity ${\bm v}({\bm r},t)$, the temperature
$T({\bm r},t)$,
the momentum flux density tensor $\underline{\Pi}({\bm r},t)$
in the reference frame moving with the local velocity ${\bm v}$,
and the heat flux density ${\bm j}_{Q}({\bm r},t)$.
They are defined by
\begin{subequations}
\label{thermoQ}
\begin{equation}
n({\bm r},t)=\!\int\!\frac{{\rm d}^{3}p}{(2\pi\hbar)^{3}}f({\bm p},{\bm r},t) \, ,
\label{n}
\end{equation}
\begin{equation}
{\bm v}({\bm r},t)=\frac{1}{n({\bm r},t)}
\!\int\!\frac{{\rm d}^{3}p}{(2\pi\hbar)^{3}}\frac{{\bm p}}{m}
f({\bm p},{\bm r},t) \, ,
\label{v}
\end{equation}
\begin{equation}
T({\bm r},t)=\frac{2}{3k_{\rm B}n({\bm r},t)}
\!\int\!\frac{{\rm d}^{3}p}{(2\pi\hbar)^{3}}\frac{\bar{p}^{2}}{2m}
f({\bm p},{\bm r},t) \, ,
\label{T}
\end{equation}
\begin{equation}
\underline{\Pi}({\bm r},t)=
\!\int\!\frac{{\rm d}^{3}p}{(2\pi\hbar)^{3}}\frac{\bar{\bm p}\bar{\bm p}}{m}
f({\bm p},{\bm r},t) \, ,
\label{Pi}
\end{equation}
\begin{equation}
{\bm j}_{Q}({\bm r},t)=
\!\int\!\frac{{\rm d}^{3}p}{(2\pi\hbar)^{3}}\frac{\bar{p}^{2}}{2m}
\frac{\bar{\bm p}}{m}f({\bm p},{\bm r},t) \, ,
\label{j_Q}
\end{equation}
\end{subequations}
with $\bar{\bm p}\!\equiv\!{\bm p}-m{\bm v}$.
The expressions (\ref{n})-(\ref{j_Q}) also result
from eqs.\ (C$\cdot$1a), (C$\cdot$1b), 
(C$\cdot$20a), (C$\cdot$14) and (C$\cdot$20b) of ref.\ \onlinecite{Kita06},
respectively, by noting eq.\ (\ref{f-def}), neglecting the
interaction terms, and identifying the local internal-energy density 
$\tilde{\cal E}$ as $\tilde{\cal E}\!=\!\frac{3}{2}nk_{\rm B}T$.

To obtain the number, momentum and energy conservation laws,
let us multiply eq.\ (\ref{Boltzmann}) by $1$, ${\bm p}$ and
$\bar{p}^{2}/2m$, respectively, 
and perform integration over ${\bm p}$.
The contribution from the collision integral (\ref{C1}) vanishes
in all the three cases
due to the particle, momentum and energy conservations 
through the collision.\cite{CC90,HCB54}
The resulting hydrodynamic equations can be written 
in terms of the quantities of eq.\ (\ref{thermoQ}) as
\begin{subequations}
\label{continuity}
\begin{equation}
\frac{\partial n}{\partial t}+{\bm \nabla}(n{\bm v})
=0 \, ,
\label{continuity-n}
\end{equation}
\begin{equation}
\frac{\partial {\bm v}}{\partial t}+{\bm v}\!\cdot\!{\bm \nabla}{\bm v}
+\frac{1}{mn}{\bm \nabla}\underline{\Pi}+g{\bm e}_{z} ={\bm 0}\, ,
\label{continuity-p}
\end{equation}
\begin{equation}
\frac{3}{2}nk_{\rm B}\!\left(\frac{\partial T}{\partial t}+
{\bm v}\cdot{\bm\nabla}T\right)\!+{\bm\nabla}\!\cdot{\bm j}_{Q}
+\underline{\Pi}:\!{\bm \nabla}{\bm v}=0 \, ,
\label{continuity-E}
\end{equation}
\end{subequations}
where ${\bm e}_{z}$ is the unit vector along the $z$ axis
and $\underline{A}\!:\!\underline{B}$ denotes 
the tensor product:
$\underline{A}\!:\!\underline{B}\!\equiv\!\sum_{ij}A_{ij}B_{ji}$.
These equations are identical in form with eqs.\ (C$\cdot$2),
(C$\cdot$9) and (C$\cdot$21) of ref.\ \onlinecite{Kita06}, respectively, 
with $U\!=\!mgz$ and $\tilde{\cal E}\!=\!\frac{3}{2}nk_{\rm B}T$.

\subsection{\label{subsec:Enskog}The Enskog series}

We now reduce the whole procedures of
solving eq.\ (\ref{Boltzmann}) 
for $({\bm p},{\bm r},t)$ 
to those of solving eq.\ (\ref{continuity}) for $({\bm r},t)$.
We adopt the well-known Enskog method\cite{CC90,HCB54}
for this purpose, i.e., the expansion from the local equilibrium.
We here describe the transformation
to the extent necessary for a later application
to Rayleigh-B\'enard convection.

Let us expand the distribution function formally as
\begin{equation}
f({\bm p},{\bm r},t)=f^{({\rm eq})}({\bm p},{\bm r},t)
\left[1\!+\!\varphi^{(1)}({\bm p},{\bm r},t)
+\cdots\right] ,
\label{Enskog}
\end{equation}
where $f^{({\rm eq})}$ is the local-equilibrium distribution 
given explicitly by
\begin{equation}
f^{({\rm eq})}=\frac{(2\pi\hbar)^{3}n}{(2\pi mk_{\rm B}T)^{3/2}}\exp\!\left(
-\frac{\bar{\bm p}^{2}}{2mk_{\rm B}T}\right) ,
\label{f_eq}
\end{equation}
with $\bar{\bm p}\!\equiv\!{\bm p}-m{\bm v}$.
This $f^{({\rm eq})}$ has been chosen so as
to satisfy the two conditions:\cite{CC90,HCB54} (i) the local equilibrium condition
that the collision integral vanishes; (ii)
eqs.\ (\ref{n})-(\ref{T}) by itself.
It hence follows that the higher-order corrections $\varphi^{(j)}$
$(j\!=\!1,2,\cdots)$ in eq.\ (\ref{Enskog})
should obey the constraints:
\begin{equation}
\int\frac{{\rm d}^{3}p}{(2\pi\hbar)^{3}}\bar{\bm p}^{n}f^{({\rm eq})}\varphi^{(j)}=0 
\hspace{5mm}(n\!=\!0,1,2)\, .
\label{constraint}
\end{equation}
Note that we have incorporated five space-time dependent parameters 
in $f^{({\rm eq})}$, i.e., $n$, ${\bm v}$ and $T$,
which can be determined completely with eqs.\
(\ref{continuity-n})-(\ref{continuity-E}) of conservation laws.
The remaining task here is to express the extra quantities
$\underline{\Pi}$ and ${\bm j}_{Q}$ in eq.\ (\ref{continuity}) 
as functionals of $n$, ${\bm v}$ and $T$.

Let us substitute eq.\ (\ref{Enskog}) into eq.\ (\ref{Boltzmann}), regard
the space-time differential operators on the left-hand side 
as first-order quantities, and make use of the fact that 
the collision integral (\ref{C1}) vanishes for $f^{({\rm eq})}$.
We thereby arrive at the first-order equation:
\begin{equation}
\frac{\partial f^{({\rm eq})}}{\partial t}+
\frac{{\bm p}}{m}\cdot\frac{\partial f^{({\rm eq})}}{\partial{\bm r}}
+\frac{\bar{p}_{z}g}{k_{\rm B}T}f^{({\rm eq})}={\cal C}^{(1)} \, ,
\label{Boltzmann1}
\end{equation}
where ${\cal C}^{(1)}$ is obtained from eq.\ (\ref{C2}) as
\begin{eqnarray}
&&\hspace{-10mm}
{\cal C}^{(1)}=\!\int\!\!\frac{{\rm d}^{3}q}{(2\pi\hbar)^{3}}\,\frac{q}{m}\!\int\!
{\rm d}\sigma\, f^{({\rm eq})}_{{\bm p}}f^{({\rm eq})}_{{\bm p}+{\bm q}}
\nonumber \\
&&\hspace{-0mm}
\times\!
\left[\varphi^{(1)}_{{\bm p}+({\bm q}-{\bm q}')/2}+
\varphi^{(1)}_{{\bm p}+({\bm q}+{\bm q}')/2}\!-\!\varphi^{(1)}_{{\bm p}}
\!-\!\varphi^{(1)}_{{\bm p}+{\bm q}}\right] .
\label{C^1}
\end{eqnarray}
With eq.\ (\ref{f_eq}), the derivatives of $f^{({\rm eq})}$ in eq.\ (\ref{Boltzmann1}) 
are transformed into those of $n$, ${\bm v}$ and $T$.
We then remove the time derivatives by using eq.\ (\ref{continuity})
in the local-equilibrium approximation, where 
\begin{equation}
\underline{\Pi}^{({\rm eq})}\!=\!P\underline{1}\, ,\hspace{10mm}
{\bm j}_{Q}^{({\rm eq})}\!=\!{\bm 0}\, ,
\label{j_QPi^eq}
\end{equation}
with $P\!=\!nk_{\rm B}T$ the pressure and $\underline{1}$ the unit tensor.
The left-hand side of eq.\ (\ref{Boltzmann1}) is thereby transformed into
\begin{eqnarray}
&&\hspace{-10mm}
\frac{\partial f^{({\rm eq})}}{\partial t}+
\frac{{\bm p}}{m}\cdot\frac{\partial f^{({\rm eq})}}{\partial{\bm r}}
+\frac{\bar{p}_{z}g}{k_{\rm B}T}f^{({\rm eq})}
\nonumber \\
&&\hspace{-10mm}
=f^{({\rm eq})}\!\left[2
\!\left(\bar{\bm k}\bar{\bm k}\!-\!\frac{\bar{k}^{2}}{3}\underline{1}\right)
\!:\!{\bm \nabla}{\bm v}
+\!\left(\bar{k}^{2}\!-\!\frac{5}{2}\right)\!
\frac{\bar{\bm p}}{m}\!\cdot\!{\bm \nabla}\ln T\right] ,
\label{Boltzmann1-l}
\end{eqnarray}
with $\bar{\bm k}$ a dimensionless quantity defined by
\begin{subequations}
\label{dimensionless}
\begin{equation}
\bar{\bm k}\equiv\bar{\bm p}/\sqrt{2mk_{\rm B}T} \, .
\label{k-def}
\end{equation}
It is convenient to introduce the additional dimensionless quantities:
\begin{equation}
\hat{\bm v}\equiv \sqrt{\frac{m}{2k_{\rm B}T}}{\bm v} \, ,
\hspace{10mm}\hat{\bm r}\equiv n^{1/3}{\bm r} \, ,
\label{v-rHat}
\end{equation}
\end{subequations}
and the mean-free path:
\begin{equation}
l\equiv \frac{1}{4\sqrt{2}\pi a^{2}n} \, .
\label{l}
\end{equation}
Using eqs.\ (\ref{dimensionless}) and (\ref{l}) and noting
eqs.\ (\ref{dsigma}), (\ref{f_eq}), 
(\ref{C^1}) and (\ref{Boltzmann1-l}), we can transform eq.\ (\ref{Boltzmann1})
into the dimensionless form:
\begin{eqnarray}
&&\hspace{-10mm}
\frac{2\sqrt{2}}{\sqrt{\pi}ln^{1/3}}\!\int\!\frac{{\rm d}^{3}q}{4\pi}
\!\int\!\frac{{\rm d}^{3}q'}{4\pi}\frac{\delta(q'\!-\!q)}{q}{\rm e}^{-\bar{k}^{2}
-(\bar{\bm k}+{\bm q})^{2}} 
\nonumber \\
&&\hspace{-10mm}
\times\!
\left[\varphi^{(1)}_{{\bm k}+({\bm q}-{\bm q}')/2}+
\varphi^{(1)}_{{\bm k}+({\bm q}+{\bm q}')/2}\!-\!\varphi^{(1)}_{{\bm k}}
\!-\!\varphi^{(1)}_{{\bm k}+{\bm q}}\right] 
\nonumber \\
&&\hspace{-13mm}
={\rm e}^{-\bar{k}^{2}}
\!\left[2
\!\left(\bar{\bm k}\bar{\bm k}\!-\!\frac{\bar{k}^{2}}{3}\underline{1}\right)
\!:\!\hat{\bm\nabla}\hat{\bm v}
+\!\left(\bar{k}^{2}\!-\!\frac{5}{2}\right)\!
\bar{\bm k}\!\cdot\!\hat{\bm\nabla}\ln T\right] 
,
\label{Boltzmann1b}
\end{eqnarray}
where $\hat{\bm\nabla}\!\equiv\!{\partial}/{\partial\hat{\bm r}}$,
and we have redefined $\varphi^{(1)}$ as a function of ${\bm k}
\!\equiv\!{\bm p}/\sqrt{2mk_{\rm B}T}$.
Similarly, eq.\ (\ref{constraint}) now reads
\begin{equation}
\int {\rm d}^{3}k\,{\rm e}^{-\bar{k}^{2}}\bar{\bm k}^{n}\varphi^{(1)}_{{\bm k}}
=0 \hspace{5mm} (n=0,1,2) \, .
\label{constraint2}
\end{equation}
Equation (\ref{Boltzmann1b}) with subsidiary condition (\ref{constraint2})
forms a linear integral equation for $\varphi^{(1)}_{{\bm k}}$.

The right-hand side of eq.\ (\ref{Boltzmann1b}) suggests 
that we may seek the solution in the form:\cite{CC90,HCB54}
\begin{eqnarray}
&&\hspace{-10mm}
\varphi^{(1)}_{\bm k}=-ln^{1/3}
\!\left[ A^{5/2}(\bar{k}^{2})
\left(\bar{\bm k}\bar{\bm k}-\frac{\bar{k}^{2}}{3}\underline{1}\right)
:\hat{\bm\nabla}\hat{\bm v}\right.
\nonumber \\
&&\hspace{10mm}\left.
+A^{3/2}(\bar{k}^{2})\,
\bar{\bm k}\!\cdot\!\hat{\bm\nabla}\ln T
\,\right] ,
\label{varphi}
\end{eqnarray}
where $A^{5/2}$ and $A^{3/2}$ are two unknown functions;
the use of fractions $\alpha\!=\!5/2$ and $3/2$
to distinguish them will be rationalized shortly.
Substituting eq.\ (\ref{varphi}) into it, we can transform
eq.\ (\ref{Boltzmann1b}) into 
separate equations for $A^{\alpha}$ as
\begin{eqnarray}
&&\hspace{-6mm}
\frac{2\sqrt{2}}{\sqrt{\pi}}\!\int\!\frac{{\rm d}^{3}q}{4\pi}
\!\int\!\frac{{\rm d}^{3}q'}{4\pi}\,\frac{\delta(q'\!-\!q)}{q}
{\rm e}^{-\bar{k}^{2}
-(\bar{\bm k}+{\bm q})^{2}} \left[\,{\cal T}^{\alpha}_{\bar{\bm k}}+
{\cal T}^{\alpha}_{\bar{\bm k}+{\bm q}}
\right.
\nonumber \\
&&\hspace{-6mm}
\left.
-{\cal T}^{\alpha}_{\bar{\bm k}+({\bm q}-{\bm q}')/2}
-{\cal T}^{\alpha}_{\bar{\bm k}+({\bm q}+{\bm q}')/2}\right] =
{\rm e}^{-\bar{k}^{2}}{\cal R}_{\bar{\bm k}}^{\alpha}\, ,
\label{Boltzmann1c}
\end{eqnarray}
where tensor ${\cal R}^{\alpha}_{\bm k}$ and ${\cal T}^{\alpha}_{\bm k}$ are 
defined in Table
\ref{tab:table1} together with another tensor ${\cal W}^{\alpha}_{\bm k}$.
\begin{table}[b]
\caption{\label{tab:table1}Quantities appearing in eq.\ (\ref{Boltzmann1c}).
Here $S_{0}^{\alpha}(\varepsilon)\!=\!1$ and 
$S_{1}^{\alpha}(\varepsilon)\!=\!1\!+\!\alpha\!-\!\varepsilon$.}
\vspace{1mm}
\begin{tabular}{cccc}
\vspace{1mm}
$\alpha$ & ${\cal W}^{\alpha}_{\bm k}$ & ${\cal R}^{\alpha}_{\bm k}$ & 
${\cal T}^{\alpha}_{\bm k}$ \\
\vspace{1mm}
$5/2$ 
& $\displaystyle{\bm k}{\bm k}\!-\!({k}^{2}/{3})\underline{1}$ 
& $2S_{0}^{\alpha}({k}^{2}){\cal W}^{\alpha}_{\bm k}$ 
& $A^{\alpha}({k}^{2}){\cal W}^{\alpha}_{\bm k}$ \\
\vspace{1mm}
$3/2$  & ${\bm k}$ & 
$-S_{1}^{\alpha}({k}^{2}){\cal W}^{\alpha}_{\bm k}$ 
& $A^{\alpha}({k}^{2}){\cal W}^{\alpha}_{\bm k}$
\end{tabular}   
\end{table}
Since the factor ${\rm e}^{-\bar{k}^{2}}$ is present on the right-hand side
of eq.\ (\ref{Boltzmann1c}), we expand
$A^{\alpha}(\varepsilon)$ further
in the Sonine polynomials
$S_{\ell}^{\alpha}(\varepsilon)$ as\cite{CC90,HCB54} 
\begin{equation}
A^{\alpha}(\varepsilon)=\sum_{\ell=0}^{\infty}c_{\ell}^{\alpha}S_{\ell}^{\alpha}(\varepsilon) \, .
\label{A-exp}
\end{equation}
Use of two different complete sets
$\{S_{\ell}^{\alpha}\}_{\ell}$ ($\alpha\!=\!5/2,3/2$)
is only for convenience to transform the right-hand side 
of eq.\ (\ref{Boltzmann1c}) into a vector with a single nonzero
element.
With eq.\ (\ref{varphi}) and (\ref{A-exp}) and
the orthogonality of $S_{\ell}^{\alpha}(\varepsilon)$,
we find that the constraint (\ref{constraint2})
reduces to the single condition:
\begin{equation}
c_{0}^{3/2}=0 \, .
\label{constraint3}
\end{equation}
We hence remove the $\ell\!=\!0$ term of $\alpha\!=\!3/2$
from eq.\ (\ref{A-exp}) in the subsequent discussion.
We now take the tensor ($\alpha\!=\!5/2$) or the vector ($\alpha\!=\!3/2$)
product of eq.\ (\ref{Boltzmann1c}) with
$S_{\ell}^{\alpha}(\bar{k}^{2}){\cal W}^{\alpha}_{\bar{\bm k}}/4\pi$
and perform integration over $\bar{\bf k}$.
Equation (\ref{Boltzmann1c}) is thereby transformed into
an algebraic equation for the expansion coefficients $\{c_{\ell}^{\alpha}\}_{\ell}$ as 
\begin{equation}
\sum_{\ell'}{\cal T}_{\ell\ell'}^{\alpha}c_{\ell'}^{\alpha}={\cal R}_{\ell}^{\alpha} \, .
\label{Boltzmann1d}
\end{equation}
Here ${\cal R}_{\ell}^{\alpha}$ is defined by
\begin{eqnarray}
&&\hspace{-10mm}
{\cal R}_{\ell}^{\alpha}\equiv \int\frac{{\rm d}^{3}k}{4\pi}{\rm e}^{-{k}^{2}}
S_{\ell}^{\alpha}({k}^{2})\,({\cal W}_{\bm k}^{\alpha},{\cal R}_{\bm k}^{\alpha})
\nonumber \\
&&\hspace{-4mm}
=\left\{
\begin{array}{ll}
\vspace{2mm}
\displaystyle \frac{5}{4}\sqrt{\pi}\,\delta_{\ell 0} & :\alpha=5/2 \\
\displaystyle -\frac{15}{16}\sqrt{\pi}\,\delta_{\ell 1} & :\alpha=3/2
\end{array}
\right. \! ,
\end{eqnarray}
with $({\cal W},{\cal R})\!\equiv\!{\cal W}\!:\!{\cal R}$ and
${\cal W}\!\cdot\!{\cal R}$ for $\alpha\!=\!5/2$
and $3/2$, respectively.
Also, ${\cal T}_{\ell\ell'}^{\alpha}$ is obtained with a change of variable
$\bar{\bm k}\!\rightarrow\!{\bm k}\!-\!{\bm q}/2$ in the $\bar{\bm k}$
integral as
\begin{eqnarray}
&&\hspace{-6mm}
{\cal T}_{\ell\ell'}^{\alpha}
\nonumber \\
&& \hspace{-6mm}
= \frac{2\sqrt{2}}{\sqrt{\pi}}\int_{0}^{\infty}\!{\rm d}k\, 
{\rm e}^{-2k^{2}}k^{2}
\int_{0}^{\infty}\!{\rm d}q\, {\rm e}^{-q^{2}/2}q^{3}
\int_{0}^{\infty}\!{\rm d}q'\,\delta(q'\!-\! q)
\nonumber \\
&& \hspace{-2mm}
\times \bigl[I_{\ell\ell'}^{\alpha}(k,q,q)
\!+\!I_{\ell\ell'}^{\alpha}(k,q,-q)
-2I_{\ell\ell'}^{\alpha}(k,q,q')\bigr] \, ,
\label{T_nn'}
\end{eqnarray}
with $I_{\ell\ell'}^{\alpha}(k,q,q')$ defined by
\begin{eqnarray}
&&\hspace{-6mm}
I_{\ell\ell'}^{\alpha}(k,q,q')
\nonumber \\
&&\hspace{-6mm}
\equiv 
\int \frac{{\rm d}\Omega_{\bf q}}{4\pi}
\!\int \frac{{\rm d}\Omega_{{\bf q}'}}{4\pi}
S^{\alpha}_{\ell}(({\bf k}\!-\!{\bf q}/2)^{2})
S^{\alpha}_{\ell'}(({\bf k}\!-\!{\bf q}'/2)^{2})
\nonumber \\
&&\hspace{-2mm}
\times({\cal W}_{{\bf k}-{\bf q}/2}^{\alpha},
{\cal W}_{{\bf k}-{\bf q}'/2}^{\alpha}) \, .
\label{I_pp'}
\end{eqnarray}
The quantities $I_{\ell\ell'}^{\alpha}(k,q,q)$ and $I_{\ell\ell'}^{\alpha}(k,q,-q)$
are obtained from eq.\ (\ref{I_pp'}) by removing the integral over
${\rm d}\Omega_{{\bf q}'}/4\pi$ 
and setting ${\bf q}'\!\rightarrow\!{\bf q}$ and $
{\bf q}'\!\rightarrow\!-{\bf q}$ in the integrand,
respectively.

The first few series of eq.\ (\ref{T_nn'}) are easily calculated
analytically as 
${\cal T}_{00}^{5/2}\!=\!{\cal T}_{11}^{3/2}\!=\! 1$, 
${\cal T}_{01}^{5/2}\!=\!{\cal T}_{12}^{3/2}\!=\! -1/4$, 
${\cal T}_{11}^{5/2}\!=\! 205/48$ and ${\cal T}_{22}^{3/2}\!=\! 45/16$, 
in agreement with the values given below eq.\ (10.21,3) 
of Chapman and Cowling.\cite{CC90}
The matrix element for a general $\ell\ell'$ can evaluated numerically.
With ${\cal T}_{\ell\ell'}^{\alpha}$ and ${\cal R}_{\ell}^{\alpha}$ thus
obtained, eq.\ (\ref{Boltzmann1d}) is solved
by cutting the infinite series at a finite value $\ell_{\rm c}$,
and  $\ell_{\rm c}$ is increased subsequently to check the convergence.
Table \ref{tab:table2} lists the values of $c_{\ell}^{\alpha}$
thereby obtained.
Those of $c_{0}^{5/2}$ and $c_{1}^{3/2}$ are about $2$\% larger in magnitude
than the analytic ones $5\sqrt{\pi}/4$ and $-15\sqrt{\pi}/16$
with $\ell_{\rm c}\!=\! 0$ and $1$
for $\alpha\!=\!5/2$ and $3/2$, respectively.
This rapid convergence as a function of $\ell_{\rm c}$ was already pointed out
by Chapman and Cowling.\cite{CC90}
\begin{table}[b]
\caption{\label{tab:table2}Values of $c_{\ell}^{\alpha}$ obtained by solving eq.\
(\ref{Boltzmann1d}).}
\vspace{1mm}
\begin{tabular}{cccccc}
\vspace{1mm}
$\alpha$ & $c_{0}^{\alpha}$ & $c_{1}^{\alpha}$ & $c_{2}^{\alpha}$ & $c_{3}^{\alpha}$
 & $c_{4}^{\alpha}$ \\
\vspace{1mm}
$5/2$  & $2.2511$ & $0.1390$ & $0.0233$ & $0.0058$ & $0.0018$ \\
\vspace{1mm}
$3/2$  & $0$ & $-1.7036$ & $-0.1626$ & $-0.0371$ & $-0.0117$
\end{tabular}   
\end{table}

Substituting eqs.\ (\ref{Enskog}), (\ref{f_eq}) and (\ref{varphi}) into eqs.\
(\ref{Pi}) and (\ref{j_Q}), we arrive at the first-order contributions to 
the momentum flux density tensor and the thermal flux density as
\begin{subequations}
\label{j_qPi^1}
\begin{equation}
{\Pi}_{ij}^{(1)}=-mn\nu\!\left(\frac{\partial v_{i}}{\partial r_{j}}\!+\!
\frac{\partial v_{j}}{\partial r_{i}}-\delta_{ij}\frac{2}{3}
{\bm\nabla}\!\cdot\!{\bm v}
\right) ,
\end{equation}
\begin{equation}
{\bm j}_{Q}^{(1)}=-\frac{3}{2}nk_{\rm B}\kappa\frac{\partial T}{\partial {\bm r}} \, ,
\label{j_Q^1}
\end{equation}
\end{subequations}
respectively, where $\nu$ and  $\kappa$ are the
kinematic viscosity and the thermal diffusivity,\cite{Chandrasekhar61,CH93} 
respectively, defined by
\begin{subequations}
\label{kappa-nu}
\begin{equation}
\nu=\frac{1}{4}l\sqrt{\frac{2k_{\rm B}T}{m}}c_{0}^{5/2} \, ,
\end{equation}
\begin{equation}
\kappa=-\frac{5}{6}l\sqrt{\frac{2k_{\rm B}T}{m}}c_{1}^{3/2} \, .
\label{kappa}
\end{equation}
\end{subequations}
These quantities clearly have the same dimension.
Using them as well as the specific heat at constant pressure 
$C_{P}$ and constant volume $C_{V}$, we can introduce an important
dimensionless quantity $Pr\!=\!(\nu/\kappa)(C_{P}/C_{V})$
called the Prandtl number.\cite{CC90}
Adopting $C_{P}/C_{V}\!=\!5/3$ of the ideal monatomic gas, 
we find $Pr\!=\!0.66$
from eq.\ (\ref{kappa-nu}) and Table \ref{tab:table2},
which is in excellent agreement with the value $0.67$ 
for Ar at $T\!=\!273$K.\cite{HCB54}
Table \ref{tab:table3} lists values of relevant thermodynamic and transport
coefficients around room temperature at $1$\,atm for Ne, Ar and air.

Thus, we have successfully expressed 
$\underline{\Pi}$ and ${\bm j}_{Q}$ in terms of
$n$, ${\bm v}$ and $T$ as eqs.\ (\ref{j_QPi^eq}), (\ref{j_qPi^1}) and
(\ref{kappa-nu}) within the first-order gradient expansion.
Now, eq.\ (\ref{continuity}) with
eqs.\ (\ref{j_QPi^eq}), (\ref{j_qPi^1}) and 
(\ref{kappa-nu}) forms a closed set of
equations for the five parameters $n$, ${\bm v}$ and $T$ incorporated
in $f^{\rm eq}({\bm p},{\bm r},t)$.
After solving them, we can obtain
the distribution function $f({\bm p},{\bm r},t)$
by eqs.\ (\ref{Enskog}), (\ref{f_eq}), (\ref{varphi}), (\ref{A-exp})
and Table \ref{tab:table2},
and subsequently calculate entropy by eq.\ (\ref{S}).

\begin{table}[b]
\caption{\label{tab:table3}Values of relevant thermodynamic 
and transport coefficients under
atmospheric pressure given in CGS units 
[$\alpha\!\equiv\! V^{-1}({\partial V}/{\partial T})$].}
\begin{center}
\begin{tabular}{cccc}
 & Ne ($0^{\circ}$C) & Ar ($0^{\circ}$C) & air ($0^{\circ}$C)
\\
$mn$ & $0.900\!\times\! 10^{-3}$& $1.78\!\times\! 10^{-3}$
&$1.29\!\times\! 10^{-3}$
\\
$\alpha$ & $3.66\!\times\! 10^{-3}$ & $3.67\!\times\! 10^{-3}$
& $3.67\!\times\! 10^{-3}$
\\
$\nu$  & $33.0\!\times\! 10^{-2}$& $11.8\!\times\! 10^{-2}$
& $13.2\!\times\! 10^{-2}$
\\
$\kappa$ & $83.1\!\times\! 10^{-2}$ & $29.2\!\times\! 10^{-2}$
& $25.9\!\times\! 10^{-2}$
\end{tabular}
\end{center}
\end{table}

\section{\label{sec:RB}Application to Rayleigh-B\'enard convection}

We now apply the equations of \S\ref{sec:EqMot} for $n$, ${\bm v}$ and $T$ to 
Rayleigh-B\'enard convection of a dilute classical monatomic gas
confined in the region $-d/2\!\leq\! z\!\leq \!d/2$
and $-L/2\!\leq\! x,y\!\leq\!L/2$.
The gas is heated from below so that
\begin{equation}
T(x,y,z\!=\!\pm d/2)=T_{0}\mp \Delta T/2\, ,\hspace{5mm}
\Delta T>0\, .
\label{T-boundary}
\end{equation}
The thickness $d$ and the lateral width $L$ are chosen as $l \!\ll\! d \!\ll\! L$.
It hence follows that (i) there are enough collisions along $z$ and (ii)
any effects from the side walls may be neglected.
We eventually impose the periodic boundary condition in the $xy$ plane.

We study this system by fixing 
the total particle number, total energy, and total heat flux
through $z\!=\!- d/2$.
This is equivalent to choosing
the average particle density $\bar{n}$, the average energy density $\bar{\cal E}$,
and the average heat flux density
$\bar{j}_{Q}$ at $z\!=\!-d/2$ as independent variables; hence
$T_{0}\!=\!T_{0}(\bar{n},\bar{\cal E},\bar{j}_{Q})$ and 
$\Delta T\!=\!\Delta T(\bar{n},\bar{\cal E},\bar{j}_{Q})$ 
in eq.\ (\ref{T-boundary}).
The latter two conditions also imply,
due to the energy conservation law, that there is average 
energy flux density $\bar{j}_{Q}$
through any cross section perpendicular to $z$.
The fact justifies our choice of $\bar{j}_{Q}$ as an independent 
variable to specify the system.
It should be noted that
this energy flow in the container may be due partly to
a macroscopic motion of the gas.

The standard theoretical treatment of Rayleigh-B\'enard convection starts
from introducing the Oberbeck-Boussinesq approximation to 
the equations for $n$, ${\bm v}$ and $T$.\cite{Chandrasekhar61}
However, this approximation 
seems not to have been justified in a widely accepted way.
We here develop a systematic approximation scheme 
for the equations in \S\ref{sec:EqMot} 
appropriate to treat Rayleigh-B\'enard convection, 
which will be shown to yield 
the equations with the Oberbeck-Boussinesq approximation
as the lowest-order approximation.
This consideration also enables us to estimate the entropy change
through the convective transition on a firm ground.

\subsection{\label{subsec:dimless}Introduction of dimensionless units}

We first introduce a characteristic temperature $\bar{T}$ defined by
\begin{equation}
\bar{T}\equiv 2\bar{\cal E}/3\bar{n}k_{\rm B} \, .
\end{equation}
We then adopt the units where the length, velocity and energy
are measured by $d$, $\sqrt{k_{\rm B}\bar{T}/m}$ and
$k_{\rm B}\bar{T}$, respectively.
Accordingly, we carry out a change of variables as
\begin{subequations}
\label{CV}
\begin{equation}
t=d\sqrt{\frac{m}{k_{\rm B}\bar{T}}}\,t',\hspace{5mm}
{\bm r}=d{\bm r}',
\end{equation}
and
\begin{equation}
n=\frac{n'}{d^{3}},\hspace{5mm}
{\bm v}=\sqrt{\frac{k_{\rm B}\bar{T}}{m}}\,{\bm v}',\hspace{5mm}
T=\bar{T}T'.
\end{equation}
\end{subequations}
Let us substitute eqs.\ (\ref{j_QPi^eq}), (\ref{j_qPi^1}) and 
(\ref{kappa-nu}) into eq.\ (\ref{continuity})
and subsequently perform the above change of variables.
We thereby obtain the dimensionless conservation laws:
\begin{subequations}
\label{continuity'}
\begin{equation}
\frac{\partial n'}{\partial t'}+{\bm\nabla}'\!\cdot(n'{\bm v}')=0 \, ,
\label{continuity'-n}
\end{equation}
\begin{eqnarray}
&&\hspace{-3mm}
n'\frac{\partial {\bm v}'}{\partial t'}
+n'{\bm v}'\!\cdot\!{\bm\nabla}'{\bm v}'
+{\bm\nabla}'P'
-\sum_{i}\nabla_{i}' \left[n'\nu' \!\left(
{\bm\nabla}'v_{i}'\!+\!
\nabla_{i}'{\bm v}'\right)\right]
\nonumber \\
&&\hspace{-3mm}
+\frac{2}{3}{\bm\nabla}'( n'\nu' 
{\bm\nabla}'\!\cdot{\bm v}')
+n'U_{g}'{\bm e}_{z}={\bm 0} \, ,
\label{Navier-Stokes}
\label{continuity'-v}
\end{eqnarray}
\begin{eqnarray}
&&\hspace{-11mm}
\frac{\partial T'}{\partial t'}
+{\bm v}'\!\cdot\!{\bm\nabla}'T'
-\frac{1}{n'}{\bm\nabla}'\!\cdot(
n'\kappa'{\bm\nabla}'T')
+\frac{2}{3}T'{\bm\nabla}'\!\cdot{\bm v}'
\nonumber \\
&&\hspace{-11mm}
-\frac{2\nu'}{3} 
\left[\sum_{ij}\frac{1}{2}\!\left(\frac{\partial v_{i}'}{\partial r_{j}'}+
\frac{\partial v_{j}'}{\partial r_{i}'}\right)^{\!\!2}-\frac{2}{3}\!\left(
{\bm\nabla}'\!\cdot{\bm v}'\right)^{2}\,\right]\! =0 \, ,
\label{continuity'-T}
\end{eqnarray}
\end{subequations}
with $P'\!=\!n'T'$ and 
\begin{equation}
\nu'\equiv \frac{\nu}{d}\sqrt{\frac{m}{k_{\rm B}\bar{T}}}\, ,\hspace{5mm}
\kappa'\equiv \frac{\kappa}{d}\sqrt{\frac{m}{k_{\rm B}\bar{T}}}\, ,\hspace{5mm}
U_{g}'\equiv \frac{mgd}{k_{\rm B}\bar{T}}\, .
\label{parameters}
\end{equation}
An important dimensionless quantity of the system is 
the Rayleigh number $R$ defined by
\begin{equation}
R\equiv \frac{U_{g}'\Delta T'}{\nu'\kappa'} =\frac{g\bar{T}^{-1}\Delta T d^{3}}
{\nu\kappa} \, ,
\label{Rayleigh}
\end{equation}
where $\bar{T}^{-1}$ appears as the thermal expansion coefficient $\alpha$
of the ideal gas.

The above equations will be solved by
fixing $\bar{n}$, $\bar{\cal E}\!=\!3\bar{n}k_{\rm B}\bar{T}/2$ and
$\bar{j}_{Q}$, as already mentioned.
These conditions are expressed in the dimensionless form as
\begin{subequations}
\label{nEj_Q-conditions}
\label{conditions}
\begin{equation}
\frac{1}{L^{\prime 2}}\int n'({\bm r}') \, {\rm d}^{3}r'=\bar{n}' \, ,
\label{n-condition}
\end{equation}
\begin{equation}
\frac{1}{L^{\prime 2}}\int \!
\left(\frac{3}{2}P'
+\frac{1}{2}n'v^{\prime 2}+n'U_{g}'z'\right)
{\rm d}^{3}r'=\frac{3}{2}\bar{n}'\, ,
\label{E-condition}
\end{equation}
\begin{equation}
\left.
-\frac{1}{L^{\prime 2}}\int {\rm d}x'
\int {\rm d}y'\,
\frac{3}{2}n'\kappa'\frac{\partial T'}{\partial z'}\right|_{z'=-\frac{1}{2}}
=\bar{j}_{Q}'\, ,
\label{j_Q-condition}
\end{equation}
\end{subequations}
where $P'\!=\!n'T'$, and integrations extend over $-L'/2\!\leq\! x',y'\!\leq\! L'/2$
and $-1/2\!\leq\!z'\!\leq\!1/2$.
Equation (\ref{E-condition}) has been obtained by
integration of $(p^{2}/2m\!+\!mgz)f$ over ${\bm r}$ and ${\bm p}$
with eq.\ (\ref{thermoQ}), whereas eq.\ (\ref{j_Q-condition})
originates from eq.\ (\ref{j_Q^1}).

To make an order-of-magnitude estimate for the parameters in 
eqs.\ (\ref{parameters}) and (\ref{Rayleigh}),
consider Ar of $273$K at $1$\,atm confined in a horizontal space of $d$\,cm
with the temperature difference $\Delta T$\,K.
Using Table \ref{tab:table2},
we then obtain the numbers:
\begin{subequations}
\label{parameters-Ar}
\begin{equation}
\begin{array}{ll}
\vspace{2mm}
\nu'=4.95\times 10^{-6}/d, &
\kappa'=1.22\times 10^{-5}/d, \\
U_{g}'=1.73\times 10^{-6}d, &
\Delta T'=3.67\times 10^{-3}\Delta T,
\end{array}
\end{equation}
and 
\begin{equation}
R=1.04\times 10^{2} d^{3}\Delta T .
\end{equation}
\end{subequations}
The critical Rayleigh number $R_{\rm c}$ 
for the convective transition is of the order $10^{3}$,\cite{Chandrasekhar61}
which is realized for $d\!\sim\! 2$cm and $\Delta T\!\sim\! 1$K.
We now observe that the dimensionless parameters 
have the following orders of magnitude in terms of $\delta\!\equiv\! 10^{-3}$:
\begin{equation}
\nu',\kappa',U_{g}'\sim \delta^{2},\hspace{5mm}
\Delta T'\sim \delta,\hspace{5mm}
R\sim \delta^{-1} .
\label{orders}
\end{equation}
Thus, Rayleigh-B\'enard convection is a phenomenon where
two orders of magnitude (i.e., $\delta$ and $\delta^{2}$) are relevant.
From now on we shall drop primes in every quantity of eqs.\ (\ref{continuity'})
and (\ref{conditions}).

\subsection{\label{subsec:Omit-n}Omission of the number conservation law}

Let us write eq.\ (\ref{continuity'}) in terms of
${\bm j}\!=\!n{\bm v}$ instead of ${\bm v}$.
It follows from the vector analysis that the vector field
${\bm j}$ can be written generally as
${\bm j}\!=\!{\bm \nabla}\Phi
\!+\!{\bm \nabla}\!\times\!{\bm A}$,
where $\Phi$ and ${\bm A}$ corresponds to the scalar and vector potentials
of the electromagnetic fields, respectively.
We then focus in the following only on those phenomena where
the current density satisfies
${\bm \nabla}\Phi\!=\!{\bm 0}$, i.e.,
\begin{equation}
{\bm \nabla}\!\cdot{\bm j}\!=\!{\bm 0}\, .
\label{div-j=0}
\end{equation}
This implies that we may drop eq.\
(\ref{continuity'-n}) from eq.\ (\ref{continuity'})
to treat only eqs.\ (\ref{continuity'-v}) and 
(\ref{continuity'-T}).

\subsection{\label{subsec:PowCount}Expansion in $\delta$}

Equation (\ref{orders}) suggests that we may solve
eqs.\ (\ref{continuity'-v}) and (\ref{continuity'-T})
in powers of $\delta$. 
Noting eqs.\ (\ref{n-condition}) and (\ref{E-condition}), 
we first expand $n$ and $T$ as
\begin{subequations}
\label{PowerCount}
\begin{equation}
n=\bar{n}\!\left(1+\sum_{\ell=1}^{\infty}\hat{n}^{(\ell)}\right) ,\hspace{5mm}
T=1+\sum_{\ell=1}^{\infty}T^{(\ell)} .
\label{PowerCount1}
\end{equation}
With eqs.\ (\ref{kappa-nu}), (\ref{parameters}) and (\ref{orders}),
we next expand dimensionless parameters 
$\nu\!=\!(l/4)\sqrt{2T}c_{0}^{5/2}$, $\kappa\!=\!-(5l/6)\sqrt{2T}c_{1}^{3/2}$ and 
$U_{g}$ as
\begin{equation}
\nu=\sum_{\ell=2}^{\infty}\nu^{(\ell)} ,\hspace{5mm}
\kappa=
\sum_{\ell=2}^{\infty}\kappa^{(\ell)} ,\hspace{5mm}
U_{g}=U_{g}^{(2)},
\label{PowerCount2}
\end{equation}
where $\nu^{(2)}\!=\!(l/4)\sqrt{2}c_{0}^{5/2}$ and $\kappa^{(2)}
\!=\!-(5l/6)\sqrt{2}c_{1}^{3/2}$ are constants with $l\!=\!l(\bar{n})$.
It also follows from $\Delta T\!\sim\! \delta$
and eq.\ (\ref{j_Q-condition}) that 
\begin{equation}
\bar{j}_{Q}\!=\!\bar{j}_{Q}^{(3)} \, .
\label{PowerCount3}
\end{equation}
It remains to attach orders of magnitude to the differential operators 
and ${\bf j}\!=\!n{\bm v}$.
In this context, we notice
that the Oberbeck-Boussinesq approximation yields
a critical Rayleigh number $R_{\rm c}$ which is in good 
quantitative agreement with experiment.\cite{Koschmieder93}
The fact tells us that 
the procedure to attach the orders should be carried out so
as to reproduce $R_{\rm c}$ of the Oberbeck-Boussinesq approximation.
The requirement yields
\begin{equation}
{\bm j}=\bar{n}\sum_{\ell=1}^{\infty}\hat{\bm j}^{(\ell+1/2)} \, ,\hspace{3mm}
\frac{\partial}{\partial t}=O(\delta^{1.5})\, ,\hspace{3mm}
{\bm\nabla}=O(\delta^{-0.25}) \, .
\label{PowerCount4}
\end{equation}
\end{subequations}
See eq.\ (\ref{R_c}) below and the subsequent comments for details.
The above power-counting scheme will be shown to provide not only a
justification of the Oberbeck-Boussinesq approximation
but also a systematic treatment to go beyond it.

Let us substitute eq.\ (\ref{PowerCount}) to eqs.\ (\ref{continuity'-v}) and 
(\ref{E-condition}).
The contributions of $O(\delta)$ in these equations
read $\nabla P^{(1)}\!=\!{\bm 0}$
and $\int P^{(1)}{\rm d}^{3}r\!=\!0$, respectively, 
with $P^{(1)}\!=\!\bar{n}(\hat{n}^{(1)}\!+\! T^{(1)})$.
We hence conclude $P^{(1)}\!=\! 0$, i.e.,
\begin{equation}
\hat{n}^{(1)}=-T^{(1)} \, .
\label{n^1-T^1}
\end{equation}
It also follows from eqs.\ (\ref{n-condition}) and
(\ref{j_Q-condition}) with eqs.\ (\ref{PowerCount}) and (\ref{n^1-T^1})
that $T^{(1)}$ should obey
\begin{subequations}
\label{T^1-constraint}
\begin{equation}
\int T^{(1)}\,{\rm d}^{3}r=0 \, ,
\label{T^1-constraint1}
\end{equation}
\begin{equation}
\frac{1}{L^{2}}\int_{-L/2}^{L/2}{\rm d}x
\int_{-L/2}^{L/2}{\rm d}y \left.
\frac{\partial T^{(1)}}{\partial z}\right|_{z=-1/2}=
-\frac{2\bar{j}_{Q}^{(3)}}{3\bar{n}\kappa^{(2)}} \, .
\label{T^1-constraint2}
\end{equation}
\end{subequations}
Equation (\ref{T^1-constraint}) is still not sufficient to
determine $T^{(1)}$.
It turns out below that the required equation
results from the $O(\delta^{3})$ and $O(\delta^{2.5})$
contributions 
of eqs.\ (\ref{continuity'-v}) and (\ref{continuity'-T}),
respectively.

Next, collecting terms of $O(\delta^{2})$ in eqs.\ (\ref{continuity'-v})
and (\ref{E-condition})
yield $\nabla P^{(2)}\!+\!\bar{n}U_{g}^{(2)}{\bm e}_{z}\!=\!{\bm 0}$
and $\int (\frac{3}{2}P^{(2)}\!+\!\bar{n}U_{g}^{(2)}z){\rm d}^{3}r\!=\! 0$, 
respectively.
Hence $P^{(2)}\!=\!-\bar{n}U_{g}^{(2)}z$. 
Noting $P^{(2)}\!=\!\bar{n}(\hat{n}^{(2)}\!+\!\hat{n}^{(1)}T^{(1)}\!+\! T^{(2)})$ 
and using eq.\ (\ref{n^1-T^1}), we obtain
\begin{equation}
\hat{n}^{(2)}=(T^{(1)})^{2}-T^{(2)}-U_{g}^{(2)}z .
\label{n^2-T^2}
\end{equation}
It follows from eqs.\ (\ref{n-condition}) and
(\ref{j_Q-condition}) with eqs.\ (\ref{PowerCount})
and (\ref{n^2-T^2}) that $T^{(2)}$ should obey
\begin{subequations}
\label{T^2-constraint}
\begin{equation}
\int [(T^{(1)})^{2}-T^{(2)}]\,{\rm d}^{3}r=0 \, ,
\end{equation}
\begin{equation}
\int_{-L/2}^{L/2}{\rm d}x
\int_{-L/2}^{L/2}{\rm d}y \left.\!\left(
\frac{\partial T^{(2)}}{\partial z}+\frac{T^{(1)}}{2}
\frac{\partial T^{(1)}}{\partial z}\right)\right|_{z=-1/2}=0 \, .
\label{T^2-constraint2}
\end{equation}
\end{subequations}
In deriving eq.\ (\ref{T^2-constraint2}),
use has been made of $(n\kappa)^{(3)}\!=\!\bar{n}\kappa^{(2)}T^{(1)}/2$
which results from $\kappa\!\propto\! lT^{1/2}$ and $l\!\propto\! n^{-1}$;
see eqs.\ (\ref{l}) and (\ref{kappa}).
Equation (\ref{T^2-constraint}) forms constraints on the higher-order 
contribution $T^{(2)}$, which will be irrelevant
in the present study, however.

Finally, we collect terms of $O(\delta^{3})$ in eq.\ (\ref{continuity'-v})
to obtain
\begin{eqnarray}
&&\hspace{-10mm}
\frac{\partial\hat{\bm j}^{(1.5)}}{\partial t}+
\hat{\bm j}^{(1.5)}\cdot{\bm\nabla}\hat{\bm j}^{(1.5)}
+\frac{{\bm\nabla}P^{(3)}}{\bar{n}}
\nonumber \\
&&\hspace{-10mm}
-\nu^{(2)}\nabla^{2}\hat{\bm j}^{(1.5)}+\hat{n}^{(1)}U_{g}^{(2)}{\bm e}_{z}={\bm 0}\, ,
\label{P^3}
\end{eqnarray}
where we have used eq.\ (\ref{div-j=0}).
We further operate ${\bm\nabla}\!\times\!{\bm\nabla}\times$
to the above equation and substitute eq.\ (\ref{n^1-T^1}). 
This yields
\begin{subequations}
\label{Boussinesq}
\begin{eqnarray}
&&\hspace{-11mm}
-\frac{\partial}{\partial t}\nabla^{2}\hat{\bm j}^{(1.5)}+
{\bm\nabla}\!\times\!{\bm\nabla}\!\times(
\hat{\bm j}^{(1.5)}\!\cdot{\bm\nabla}\hat{\bm j}^{(1.5)})
\nonumber \\
&&\hspace{-11mm}
+\nu^{(2)}(\nabla^{2})^{2}\hat{\bm j}^{(1.5)}
+U_{g}^{(2)}({\bm e}_{z}\nabla^{2}\!-\!{\bm e}_{z}\!\cdot\!{\bm\nabla}{\bm\nabla})
T^{(1)}={\bm 0}\, .
\end{eqnarray}
On the other hand, terms of $O(\delta^{2.5})$ in eq.\ (\ref{continuity'-T})
lead to
\begin{equation}
\frac{\partial T^{(1)}}{\partial t}+\hat{\bm j}^{(1.5)}\cdot{\bm\nabla}T^{(1)}
-\kappa^{(2)}\nabla^{2}T^{(1)}=0 \, .
\end{equation}
\end{subequations}
Equation (\ref{Boussinesq}) forms a set of coupled differential equations for
$T^{(1)}$ and $\hat{\bm j}^{(1.5)}$,
which should be solved with eq.\ (\ref{T^1-constraint}).
It is almost identical in form with that derived
with the Oberbeck-Boussinesq approximation,
predicting the same critical Rayleigh number 
$R_{\rm c}$ as will be shown below.
The whole considerations on Rayleigh-B\'enard
convection presented in the following will be based on eq.\
(\ref{Boussinesq}) with eq.\ (\ref{T^1-constraint}).

Two comments are in order before closing the subsection. 
First, if we apply the procedure of deriving eq.\ (\ref{Boussinesq}) to
the $O(\delta^{4})$ and $O(\delta^{3.5})$ contributions
of eqs.\ (\ref{continuity'-v}) and (\ref{continuity'-T}), respectively,
we obtain coupled equations for the next-order quantities
$T^{(2)}$ and $\hat{\bm j}^{(2.5)}$, 
which should be solved with eq.\ (\ref{T^2-constraint}).
Thus, we can treat higher-order contributions systematically
in the present expansion scheme.
Second, 
eq.\ (\ref{P^3}) may be regarded as
the equation to determine $P^{(3)}$ for given $T^{(1)}$ and $\hat{\bm j}^{(1.5)}$.
It yields a relation between $T^{(3)}$ and $\hat{n}^{(3)}$,
which in turn leads to the constraint for $T^{(3)}$ 
upon substitution into eqs.\ (\ref{n-condition}) and (\ref{j_Q-condition}).
On the other hand, the equation for $T^{(3)}$ 
originates from the $O(\delta^{5})$ and $O(\delta^{4.5})$ contributions
of eqs.\ (\ref{continuity'-v}) and (\ref{continuity'-T}), respectively.
Now, one may understand the hierarchy of the approximation clearly.

\subsection{\label{subsec:entropy}Expression of entropy}

We now write down the expression of entropy in powers of $\delta$. 
Entropy of the system can be calculated by eq.\ (\ref{S}),
where the distribution function $f$ is given by
eq.\ (\ref{Enskog}) with eqs.\ (\ref{f_eq}) and (\ref{varphi}).
Hereafter we shall drop the superscript in $\varphi^{(1)}$,
which specifies the order in the gradient expansion, 
to remove possible confusion with the expansion of eq.\ (\ref{PowerCount}).
Thus, $f$ is now expressed as $f\!=\!f^{({\rm eq})}(1\!+\!\varphi)$.

We first focus on $\varphi$ and write eq.\ (\ref{varphi}) 
in the present units with noting eq.\ (\ref{dimensionless}).
We then realize that $\varphi$ is proportional to $l{\bm\nabla}T$
or $l{\bm\nabla}{\bm v}$,
which are quantities of $O(\delta^{3})$ and $O(\delta^{3.5})$
in the expansion scheme of eq.\ (\ref{PowerCount}), respectively.
It hence follows that there is no contribution of $O(\delta^{2})$
from $\varphi$.
In contrast, $f^{({\rm eq})}$ yields terms of $O(\delta^{2})$,
as seen below. 
Thus, we only need to consider $f^{({\rm eq})}$.

Let us write $f^{({\rm eq})}$ of eq.\ (\ref{f_eq}) 
in the present units, substitute eq.\ (\ref{PowerCount}) into it,
and expand the resulting expression in powers of $\delta$.
We also drop terms connected with ${\bm j}$ (i.e., ${\bm v}$)
which have vanishing contribution to $S$ within $O(\delta^{2})$ 
after the momentum integration in eq.\ (\ref{S}).
We thereby obtain the relevant expansion:
\begin{eqnarray}
&&\hspace{-11mm}
f^{({\rm eq})}=\frac{(2\pi\hbar)^{3}\bar{n}}{(2\pi)^{3/2}}{\rm e}^{-\varepsilon}
\left(1\!+\sum_{\ell=1}^{2}\hat{n}^{(\ell)}\right)
\nonumber \\
&&\hspace{1mm}\times
\left[1\!+u^{(1)}(\varepsilon)\sum_{\ell=1}^{2}T^{(\ell)}+
u^{(2)}(\varepsilon)(T^{(1)})^{2}\right] ,
\label{f_eq-exp}
\end{eqnarray}
where $\varepsilon\!=\!p^{2}/2$, and $u^{(1)}$ and $u^{(2)}$ are defined by
\begin{equation}
u^{(1)}(\varepsilon)=\varepsilon-\frac{3}{2}\, ,\hspace{5mm}
u^{(2)}(\varepsilon)=\frac{1}{2}\!\left(\varepsilon^{2}-5\varepsilon+\frac{15}{4}\right).
\end{equation}
Let us substitute eq.\ (\ref{f_eq-exp}) into eq.\ (\ref{S})
and carry out integration over ${\bf p}$.
The contribution of $O(1)$ is easily obtained as ($k_{\rm B}\!=\!1$)
\begin{equation}
S^{(0)}\!=\!\bar{n}\!\left[\ln 
\frac{(2\pi)^{3/2}}{(2\pi\hbar)^{3}\bar{n}}\!+\!\frac{5}{2}\right] ,
\end{equation}
which is just the equilibrium expression\cite{Reif} 
for density $\bar{n}$ and temperature $\bar{T}$
in the conventional units, as it should.
Next, we find $S^{(1)}\!=\!0$ due to eqs.\ (\ref{n^1-T^1}) and
(\ref{T^1-constraint1}).
Thus, the contribution characteristic of heat conduction 
starts from the second order.
A straightforward calculation yields
\begin{equation}
S^{(2)}=-\frac{5}{4{\cal V}}\int \bigl(T^{(1)}\bigr)^{2}\, {\rm d}^{3}r \, .
\label{S^2}
\end{equation}
Equation (\ref{S^2}) is the basic starting point to calculate the entropy change
through the convective transition.
Note that we have fixed $\bar{j}_{Q}$ in the present consideration,
i.e., the initial temperature slope as seen from
eq.\ (\ref{T^1-constraint2}).

At this state, 
it may be worthwhile to present a qualitative argument on entropy of
Rayleigh-B\'enard convection.
With the initial temperature slope fixed as eq.\ (\ref{T^1-constraint2}),
eq.\ (\ref{S^2}) tells us that entropy will be larger 
as the temperature profile becomes more uniform
between $z\!=\!\pm 1/2$.
The conducting state with ${\bm v}\!=\!{\bf 0}$ has the 
linear temperature profile, as shown shortly below in eq.\ (\ref{T^1-conducting}).
Thus, any increase of entropy over this conducting state
is brought about by reducing the temperature difference
between $z\!=\!\pm 1/2$.
Such a state necessarily accompanies a
temperature variation which is weaker around
$z\!=\!0$ than near the boundaries $z\!=\!\pm 1/2$.
This temperature profile is indeed an essential feature of Rayleigh-B\'enard convection 
which shows up as an increase of the Nusselt number (i.e., the efficiency of
heat transport) under fixed temperature 
difference.\cite{Chandrasekhar61,Koschmieder93}
Combining eqs.\ (\ref{T^1-constraint2}) and (\ref{S^2}) with 
the experimental observation on the Nusselt number, 
we thereby conclude without any detailed calculations
that entropy of Rayleigh-B\'enard convection should be larger than 
entropy of the conducting state in the present conditions with 
$\bar{j}_{Q}=$const.
Thus, Rayleigh-B\'enard convection is expected to satisfy 
the principle of maximum entropy given at the beginning of the paper.
We shall confirm this fact below through detailed numerical studies.

\section{\label{sec:Fourier}Periodic solution with stress-free boundaries}

Equation (\ref{Boussinesq}) with eq.\ (\ref{T^1-constraint})
forms a set of simultaneous equations for $T^{(1)}$ and $\hat{\bm j}^{(1.5)}$,
which should be supplemented by the boundary condition on
$\hat{\bm j}^{(1.5)}$.
For simplicity, we here adopt the assumption of stress-free boundaries:\cite{Chandrasekhar61}
\begin{equation}
\hat{j}_{z}^{(1.5)}=
\frac{\partial^{2}}{\partial z^{2}}\hat{j}_{z}^{(1.5)}=0 \hspace{3mm}{\rm at}
\hspace{3mm} z=\pm\frac{1}{2} \, .
\label{Stress-Free-BC}
\end{equation}
However, qualitative features of the convective solutions will
be universal among the present and more realistic/complicated boundary conditions;
see the argument at the end of the preceding section.

We first discuss the heat-conducting solution of eq.\ (\ref{Boussinesq})
and its instability towards convection.
We then transform eq.\ (\ref{Boussinesq}) with eqs.\ (\ref{T^1-constraint}) and
(\ref{Stress-Free-BC}) in a form suitable to
obtain periodic convective structures.

\subsection{\label{subsec:ConductingSol}Conducting solution}

Let us consider the conducting solution of eq.\ (\ref{Boussinesq})
where $\hat{\bm j}^{(1.5)}\!=\!{\bm 0}$
with uniformity in the $xy$ plane.
Equation (\ref{Boussinesq}) then 
reduces to ${\rm d}^{2} T^{(1)}/{\rm d}z^{2}\!=\!0$, which is solved 
with eq.\ (\ref{T^1-constraint}) as
\begin{equation}
T^{(1)}=-\Delta T_{\rm hc} z \, ,\hspace{5mm}\Delta T_{\rm hc}\equiv
\frac{2\bar{j}_{Q}^{(3)}}{3\bar{n}\kappa^{(2)}}\, .
\label{T^1-conducting}
\end{equation}
Substituting this expression into eq.\ (\ref{S^2}), we obtain
entropy of the conducting state measured from $S^{(0)}$ as
\begin{equation}
S^{(2)}_{\rm hc}=-\frac{5}{48}(\Delta T_{\rm hc})^{2}
\, .
\label{S^2-hc}
\end{equation}

\subsection{\label{subsec:instability}Instability of the conducting state}

We next check stability of the conducting solution 
by adding a small perturbation given by
\begin{subequations}
\label{stability}
\begin{equation}
\hat{\bm j}^{(1.5)}({\bm r},t)={\rm e}^{\lambda t}
{\rm e}^{i{\bm k}_{\perp}\cdot{\bm r}}(\delta{\bm j}^{\rm s}\sin k_{z}\zeta
+\delta{\bm j}^{\rm c}_{\perp}\cos k_{z}\zeta) \, ,
\end{equation}
\begin{equation}
T^{(1)}({\bm r},t)=-\Delta T_{\rm hc} z+\delta T{\rm e}^{\lambda t}
{\rm e}^{i{\bm k}_{\perp}\cdot{\bm r}}\sin k_{z}\zeta \, ,
\end{equation}
\end{subequations}
where $\zeta\equiv z\!+\!1/2$,
$k_{z}\!=\!\ell_{3}\pi$ $(\ell_{3}\!=\!1,2,\cdots)$ 
from eq.\ (\ref{Stress-Free-BC}),
and $\delta{\bm j}^{\rm c}_{\perp}$ denotes a vector in the $xy$ plane.
Let us substitute eq.\ (\ref{stability}) into eq.\ (\ref{Boussinesq}) and
linearize it with respect to the perturbation.
This leads to
\begin{subequations}
\label{linear}
\begin{equation}
(\lambda\!+\!\nu^{(2)} k^{2})k^{2}\delta{\bm j}^{\rm s}-
U_{g}^{(2)}k_{\perp}^{2}\delta T {\bm e}_{z}={\bm 0} \, ,
\label{linear1}
\end{equation}
\begin{equation}
(\lambda\!+\!\nu^{(2)} k^{2})k^{2}\delta{\bm j}^{\rm c}_{\perp}-i
U_{g}^{(2)}{\bm k}_{\perp}k_{z}\delta T ={\bm 0} \, ,
\label{linear2}
\end{equation}
\begin{equation}
(\lambda\!+\!\kappa^{(2)} k^{2})\delta T-
\Delta T_{\rm hc}\delta j_{z}^{\rm s} =0 \, .
\label{linear3}
\end{equation}
\end{subequations}
The components $\delta{\bm j}^{\rm s}_{\perp}$ and
$\delta{\bm j}^{\rm c}_{\perp}$ in the $xy$ plane 
are obtained from eqs.\ (\ref{linear1}) and 
(\ref{linear2}) as
\begin{equation}
\delta{\bm j}^{\rm s}_{\perp}={\bm 0}\, ,\hspace{5mm}
\delta{\bm j}^{\rm c}_{\perp}=\frac{i
U_{g}^{(2)}{\bm k}_{\perp}k_{z}}{(\lambda\!+\!\nu^{(2)} k^{2})k^{2}}\delta T\, .
\label{dj_perp}
\end{equation}
In contrast, the $z$ component of eq.\ (\ref{linear1}) and eq.\ (\ref{linear3})
form linear homogeneous equations for $\delta j_{z}^{\rm s}$ and $\delta T$.
The requirement that they have a non-trivial solution yields
\begin{equation}
\lambda^{2}\!+\!(\nu^{(2)}\!+\!\kappa^{(2)})k^{2}\lambda
\!+\!\nu^{(2)}\kappa^{(2)}(k^{4}\!-\!R^{(-1)}k_{\perp}^{2}/k^{2})
\!=\!0\, ,
\label{insta-eq}
\end{equation}
%, 
with $R^{(-1)}$ the Rayleigh number defined by
eq.\ (\ref{Rayleigh}) with $U_{g}'\!\rightarrow\!U_{g}^{(2)}$,
$\Delta T\!\rightarrow\!\Delta T_{\rm hc}$,
$\nu'\!\rightarrow\!\nu^{(2)}$ and
$\kappa'\!\rightarrow\!\kappa^{(2)}$. 
The conducting solution becomes
unstable when eq.\ (\ref{insta-eq}) has a positive solution, i.e.,
\begin{equation}
R^{(-1)}\equiv\!\frac{U_{g}^{(2)}\Delta T_{\rm hc}}{\kappa^{(2)}\nu^{(2)}}
\geq\frac{(k_{\perp}^{2}\!+\!k_{z}^{2})^{3}}{k_{\perp}^{2}} \geq \frac{27\pi^{4}}{4} \, .
\label{R_c}
\end{equation}
Thus, we have obtained the value $R_{\rm c}\!=\!27\pi^{4}/4$
for the critical Rayleigh number\cite{Chandrasekhar61,Busse78}
which corresponds to $(k_{\perp},k_{z})\!=\!(\pi/\sqrt{2},\pi)$.

Besides reproducing the established results,\cite{Chandrasekhar61}
the above consideration may also be important in the following respects.
First, we require that: (i) terms of
the $z$ component of eq.\ (\ref{linear1}) all have the same order in $\delta$
with $\delta T\!=\!O(\delta)$; 
(ii) the same be true for terms of eq.\ (\ref{linear3}).
This leads to the attachment of the order-of-magnitude:
$\delta{\bm j}\!=\!O(\delta^{1.5})$, $k\!=\!O(\delta^{-0.25})$
and $\lambda\!=\!O(\delta^{1.5})$,
thereby justifying eq.\ (\ref{PowerCount4}).
The conclusion $k\!=\!O(\delta^{-0.25})$ also results from 
$k\!=\!\sqrt{3/2}\pi\!\sim\!4$ for the critical Rayleigh number.
Second, eq.\ (\ref{R_c}) removes the ambiguity in $\nu$, $\kappa$
and $\alpha$ to estimate the critical Rayleigh number $R_{\rm c}$.
Specifically, we should use the mean values over $-1/2\!\leq\!z\!\leq\!1/2$
for a detailed comparison of $R_{\rm c}$ between theory and experiment.

\subsection{Convective solution}

We now focus on the convective solution of eq.\ (\ref{Boussinesq})
with periodic structures.
Let us introduce basic vectors as
\begin{equation}
{\bm a}_{1}=(a_{1x},a_{1y},0),\hspace{3mm}
{\bm a}_{2}=(0,a_{2},0),\hspace{3mm}
{\bm a}_{3}={\bm e}_{z}.
\end{equation}
We consider the region in the $xy$ plane spanned by ${\cal N}_{1}{\bm a}_{1}$
and ${\cal N}_{2}{\bm a}_{2}$ with ${\cal N}_{j}$ ($j\!=\!1,2$) a huge integer,
and impose the periodic boundary condition.
The wave vector ${\bm k}$ is then defined in terms of the reciprocal lattice vectors
${\bm b}_{1}=2\pi({\bm a}_{2}\!\times\!{\bm a}_{3})/[({\bm a}_{1}\!\times\!{\bm a}_{2})
\!\cdot\!{\bm a}_{3}]$, ${\bm b}_{2}=2\pi({\bm a}_{3}\!\times\!{\bm a}_{1})/[({\bm a}_{1}\!\times\!{\bm a}_{2})\!\cdot\!{\bm a}_{3}]$ and ${\bm b}_{3}\!=\!\pi{\bm e}_{z}$ as
\begin{equation}
{\bm k}=\sum_{j=1}^{3}\ell_{j}{\bm b}_{j} \, ,
\label{k-l}
\end{equation}
with $\ell_{j}$ denoting an integer.
The analysis of \S\ref{subsec:instability} suggests that
the stable solution satisfies
$|{\bm b}_{1}|\!\sim \!|{\bm b}_{2}|\!\sim \!\pi/\sqrt{2}$.

Equation (\ref{dj_perp}) tells us that $\delta{\bm j}_{\perp}$ and
$\delta T$ are out-of-phase. 
Also noting eqs.\ (\ref{T^1-constraint}) and (\ref{Stress-Free-BC}), 
we now write down the steady solution of eq.\ (\ref{Boussinesq}) in the form:
\begin{subequations}
\label{jT-expand}
\begin{equation}
\hat{\bm j}_{\perp}^{(1.5)}({\bm r})=
\sum_{{\bm k}_{\perp}(\neq{\bf 0})}\sum_{k_{z}}\tilde{\bm j}_{\perp{\bm k}}
\sin({\bm k}_{\perp}\!\cdot{\bm r})\cos k_{z}\zeta \, ,
\label{jT-expand1}
\end{equation}
\begin{equation}
\hat{j}_{z}^{(1.5)}({\bm r})=
\sum_{{\bm k}_{\perp}(\neq{\bf 0})}\sum_{k_{z}}
\tilde{j}_{z{\bm k}}\cos({\bm k}_{\perp}\!\cdot{\bm r})\sin k_{z}\zeta \, ,
\label{jT-expand2}
\end{equation}
\begin{equation}
T^{(1)}({\bm r})=
\sum_{{\bm k}_{\perp}}\sum_{k_{z}}\tilde{T}_{{\bm k}}\cos({\bm k}_{\perp}\!\cdot{\bm r})\sin k_{z}\zeta 
-\Delta T_{\rm cv} z
-T_{1}\, ,
\label{T-expand}
\end{equation}
\end{subequations}
where $\zeta\!\equiv\!z\!+\!1/2$, and we have chosen $\hat{j}_{z}^{(1.5)}$ and $T^{(1)}$
as even functions in the $xy$ plane without losing the generality.
The ${\bm k}$ summations in eq.\ (\ref{jT-expand}) run over
\begin{equation}
\left\{
\begin{array}{lll}
\vspace{1mm}
-\infty\!\leq\!\ell_{1}\!\leq\!-1, & 1\!\leq\!\ell_{2}\!\leq\!\infty , &
1\!\leq\!\ell_{3}\!\leq\!\infty , \\
0\!\leq\!\ell_{1}\!\leq\!\infty, & 0\!\leq\!\ell_{2}\!\leq\!\infty , &
1\!\leq\!\ell_{3}\!\leq\!\infty ,
\end{array}
\right. 
\label{l-range}
\end{equation}
which covers all the independent basis functions.
The condition (\ref{div-j=0})
is transformed into
\begin{subequations}
\label{div-j}
\begin{equation}
{\bm k}\!\cdot\!\tilde{\bm j}_{\perp{\bm k}}
\!+\!k_{z}\tilde{j}_{z{\bm k}}\!=\!0\, .
\end{equation}
Thus, $\tilde{\bm j}_{\perp{\bm k}}$ may be written generally as
\begin{equation}
\tilde{\bm j}_{\perp{\bm k}}=-\frac{{\bm k}_{\perp}k_{z}}{k_{\perp}^{2}}
\tilde{j}_{z{\bm k}}+\frac{{\bm e}_{z}\!\times\!{\bm k}_{\perp}}{k_{\perp}}
\tilde{j}_{p{\bm k}} \, .
\label{j_perp-k}
\end{equation}
\end{subequations}
The constants $\Delta T_{\rm cv}$ and $T_{1}$ in eq.\ (\ref{T-expand})
denote the temperature difference between $z\!=\!\pm 1/2$ and the average
temperature shift, respectively.
They are fixed so as to satisfy eq.\ (\ref{T^1-constraint}) as
\begin{subequations}
\label{T_01}
\begin{equation}
\Delta T_{\rm cv}=\Delta T_{\rm hc}+
\sum_{k_{z}}\tilde{T}_{{\bm k}_{\perp}={\bm 0},k_{z}}k_{z}\, ,
\label{DeltaT-cv}
\end{equation}
\begin{equation}
T_{1}=\sum_{k_{z}}\tilde{T}_{{\bm k}_{\perp}={\bm 0},k_{z}}
\frac{1\!-\!\cos k_{z}}{k_{z}} \, ,
\label{T_1}
\end{equation}
\end{subequations}
where $\Delta T_{\rm hc}$ denotes temperature difference of the conducting state
given explicitly in eq.\ (\ref{T^1-conducting}).

It follows from the energy conservation law that 
eq.\ (\ref{T^1-constraint2}) should also hold at $z\!=\!1/2$,
which leads to an alternative expression for $\Delta T_{\rm cv}$.
Subtracting it from eq.\ (\ref{DeltaT-cv}) yields the identity obeyed by
$\tilde{T}_{{\bm k}_{\perp}={\bm 0},k_{z}}$:
\begin{equation}
\sum_{k_{z}}\tilde{T}_{{\bm k}_{\perp}={\bm 0},k_{z}}
k_{z}(1\!-\!\cos k_{z})\!=\! 0\, ,
\label{check}
\end{equation}
which is useful to check the accuracy of numerical calculations.

Let us substitute eq.\ (\ref{jT-expand}) into eq.\ (\ref{Boussinesq}).
A straightforward calculation of using eq.\ (\ref{div-j}) then leads to 
coupled algebraic equations for $\tilde{T}_{{\bm k}}$, $\tilde{j}_{z{\bm k}}$ 
and $\tilde{j}_{p{\bm k}}$ as
\begin{subequations}
\label{BE15}
\begin{eqnarray}
&&\hspace{-3mm}
\kappa^{(2)} k^{2} \tilde{T}_{{\bm k}} -\Delta T_{\rm cv} \tilde{j}_{z{\bm k}}
\nonumber \\
&&\hspace{-3mm}
+\frac{1}{4}\sum_{{\bm k}'{\bm k}''}
\tilde{T}_{{\bm k}'}\biggl\{ {\bm k}\cdot\tilde{\bm j}_{{\bm k}''}
\biggl[-
\delta_{{\bm k}_{\perp}''+{\bm k}_{\perp}'{\bm k}_{\perp}}
\delta_{k_{z}''-k_{z}'k_{z}}
\nonumber \\
&&\hspace{-3mm}
+\delta_{{\bm k}_{\perp}''+{\bm k}_{\perp}'{\bm k}_{\perp}}
\delta_{k_{z}''+k_{z}'k_{z}}+\biggl(1\!-\!\frac{\delta_{{\bm k}_{\perp}{\bm 0}}}{2}
\biggr)\!\biggl(\delta_{{\bm k}_{\perp}''-{\bm k}_{\perp}'{\bm k}_{\perp}}
\delta_{k_{z}''+k_{z}'k_{z}}
\nonumber \\
&&\hspace{-3mm}
-\delta_{{\bm k}_{\perp}''-{\bm k}_{\perp}'{\bm k}_{\perp}}
\delta_{k_{z}''-k_{z}'k_{z}}
-\delta_{{\bm k}_{\perp}'-{\bm k}_{\perp}''{\bm k}_{\perp}}
\delta_{k_{z}'-k_{z}''k_{z}}\biggr)
\biggr]
\nonumber \\
&&\hspace{-3mm}
+\left({\bm k}_{\perp}\cdot\tilde{\bm j}_{\perp{\bm k}''}
-k_{z}\tilde{j}_{z{\bm k}''}\right)
\biggl[\delta_{{\bm k}_{\perp}'+{\bm k}_{\perp}''{\bm k}_{\perp}}
\delta_{k_{z}'-k_{z}''k_{z}}
\nonumber \\
&&\hspace{-3mm}
+\biggl(1\!-\!\frac{\delta_{{\bm k}_{\perp}{\bm 0}}}{2}
\biggr)\!\biggl(\delta_{{\bm k}_{\perp}'-{\bm k}_{\perp}''{\bm k}_{\perp}}
\delta_{k_{z}''-k_{z}'k_{z}}
+\delta_{{\bm k}_{\perp}''-{\bm k}_{\perp}'{\bm k}_{\perp}}
\delta_{k_{z}'-k_{z}''k_{z}}
\nonumber \\
&&\hspace{-3mm}
-\delta_{{\bm k}_{\perp}'-{\bm k}_{\perp}''{\bm k}_{\perp}}
\delta_{k_{z}''+k_{z}'k_{z}}\biggr)
\biggr]\biggr\}
=0 \, .
\label{BE15a}
\end{eqnarray}
\begin{eqnarray}
&&\hspace{-10mm}
\nu^{(2)} k^{2} \tilde{j}_{z{\bm k}}
-U_{g}^{(2)}\frac{k_{\perp}^{2}}{k^{2}} \tilde{T}_{{\bm k}}
\nonumber \\
&&\hspace{-10mm}
+\frac{1}{4}\sum_{{\bm k}'{\bm k}''}
\tilde{j}_{z{\bm k}'}\bigl[ {\bm k}\cdot\tilde{\bm j}_{{\bm k}''}(-
\delta_{{\bm k}_{\perp}''+{\bm k}_{\perp}'{\bm k}_{\perp}}
\delta_{k_{z}''-k_{z}'k_{z}}
\nonumber \\
&&\hspace{-10mm}
+\delta_{{\bm k}_{\perp}''-{\bm k}_{\perp}'{\bm k}_{\perp}}
\delta_{k_{z}''+k_{z}'k_{z}}
+\delta_{{\bm k}_{\perp}''+{\bm k}_{\perp}'{\bm k}_{\perp}}
\delta_{k_{z}''+k_{z}'k_{z}}
\nonumber \\
&&\hspace{-10mm}
-\delta_{{\bm k}_{\perp}''-{\bm k}_{\perp}'{\bm k}_{\perp}}
\delta_{k_{z}''-k_{z}'k_{z}}
-\delta_{{\bm k}_{\perp}'-{\bm k}_{\perp}''{\bm k}_{\perp}}
\delta_{k_{z}'-k_{z}''k_{z}})
\nonumber \\
&&\hspace{-10mm}
+({\bm k}_{\perp}\cdot\tilde{\bm j}_{\perp{\bm k}''}
-k_{z}\tilde{j}_{z{\bm k}''})
(\delta_{{\bm k}_{\perp}'-{\bm k}_{\perp}''{\bm k}_{\perp}}
\delta_{k_{z}''-k_{z}'k_{z}}
\nonumber \\
&&\hspace{-10mm}
+\delta_{{\bm k}_{\perp}''-{\bm k}_{\perp}'{\bm k}_{\perp}}
\delta_{k_{z}'-k_{z}''k_{z}}
+\delta_{{\bm k}_{\perp}'+{\bm k}_{\perp}''{\bm k}_{\perp}}
\delta_{k_{z}'-k_{z}''k_{z}}
\nonumber \\
&&\hspace{-10mm}
-\delta_{{\bm k}_{\perp}'-{\bm k}_{\perp}''{\bm k}_{\perp}}
\delta_{k_{z}''+k_{z}'k_{z}}
)\bigr]
\nonumber \\
&& \hspace{-10mm}
-\frac{k_{z}}{4k^{2}}\sum_{{\bm k}'{\bm k}''}
\bigl[ 
({\bm k}_{\perp}\cdot\tilde{\bm j}_{\perp{\bm k}'}
-k_{z}\tilde{j}_{z{\bm k}'}){\bm k}\cdot\tilde{\bm j}_{{\bm k}''}
\nonumber \\
&&\hspace{-10mm}
\times
(\delta_{{\bm k}_{\perp}''+{\bm k}_{\perp}'{\bm k}_{\perp}}
\delta_{k_{z}''- k_{z}'k_{z}}
-\delta_{{\bm k}_{\perp}''-{\bm k}_{\perp}'{\bm k}_{\perp}}
\delta_{k_{z}''+ k_{z}'k_{z}})
\nonumber \\
&& \hspace{-10mm}
+{\bm k}\cdot\tilde{\bm j}_{{\bm k}'}{\bm k}\cdot\tilde{\bm j}_{{\bm k}''}
(\delta_{{\bm k}_{\perp}''+{\bm k}_{\perp}'{\bm k}_{\perp}}
\delta_{k_{z}''+k_{z}'k_{z}}
\nonumber \\
&&\hspace{-10mm}
-\delta_{{\bm k}_{\perp}''-{\bm k}_{\perp}'{\bm k}_{\perp}}
\delta_{k_{z}''-k_{z}'k_{z}}
-\delta_{{\bm k}_{\perp}'-{\bm k}_{\perp}''{\bm k}_{\perp}}
\delta_{k_{z}'-k_{z}''k_{z}})
\nonumber \\
&& \hspace{-10mm}
-({\bm k}_{\perp}\cdot\tilde{\bm j}_{\perp{\bm k}'}
-k_{z}\tilde{j}_{z{\bm k}'})({\bm k}_{\perp}\cdot\tilde{\bm j}_{\perp{\bm k}''}
-k_{z}\tilde{j}_{z{\bm k}''})
\nonumber \\
&&\hspace{-10mm}
\times
(\delta_{{\bm k}_{\perp}'-{\bm k}_{\perp}''{\bm k}_{\perp}}
\delta_{k_{z}''- k_{z}'k_{z}}
+\delta_{{\bm k}_{\perp}''-{\bm k}_{\perp}'{\bm k}_{\perp}}
\delta_{k_{z}'- k_{z}''k_{z}})
\nonumber \\
&& \hspace{-10mm}
+{\bm k}\cdot\tilde{\bm j}_{{\bm k}'}
({\bm k}_{\perp}\cdot\tilde{\bm j}_{\perp{\bm k}''}
-k_{z}\tilde{j}_{z{\bm k}''})
(\delta_{{\bm k}_{\perp}'+{\bm k}_{\perp}''{\bm k}_{\perp}}
\delta_{k_{z}'- k_{z}''k_{z}}
\nonumber \\
&&\hspace{-10mm}
-\delta_{{\bm k}_{\perp}'-{\bm k}_{\perp}''{\bm k}_{\perp}}
\delta_{k_{z}'+ k_{z}''k_{z}}
)\bigr]
=0 \, ,
\label{BE15b}
\end{eqnarray}
\begin{eqnarray}
&&\hspace{-13mm}
\nu^{(2)} k^{2}\tilde{j}_{p{\bm k}}
+\frac{1}{4}\sum_{{\bm k}'{\bm k}''}
\frac{({\bm e}_{z}\!\times{\bm k}_{\perp})\cdot\tilde{\bm j}_{\perp{\bm k}'}}
{k_{\perp}}\bigl[ {\bm k}\cdot\tilde{\bm j}_{{\bm k}''}
\nonumber \\
&&\hspace{-13mm}
\times (
\delta_{{\bm k}_{\perp}''+{\bm k}_{\perp}'{\bm k}_{\perp}}
\delta_{k_{z}''-k_{z}'k_{z}}
-\delta_{{\bm k}_{\perp}''-{\bm k}_{\perp}'{\bm k}_{\perp}}
\delta_{k_{z}''+k_{z}'k_{z}}
\nonumber \\
&&\hspace{-13mm}
+\delta_{{\bm k}_{\perp}''+{\bm k}_{\perp}'{\bm k}_{\perp}}
\delta_{k_{z}''+k_{z}'k_{z}}
-\delta_{{\bm k}_{\perp}''-{\bm k}_{\perp}'{\bm k}_{\perp}}
\delta_{k_{z}''-k_{z}'k_{z}}
\nonumber \\
&&\hspace{-13mm}
-\delta_{{\bm k}_{\perp}'-{\bm k}_{\perp}''{\bm k}_{\perp}}
\delta_{k_{z}'-k_{z}''k_{z}})
-({\bm k}_{\perp}\cdot\tilde{\bm j}_{\perp{\bm k}''}
-k_{z}\tilde{j}_{z{\bm k}''})
\nonumber \\
&&\hspace{-13mm}
\times
(\delta_{{\bm k}_{\perp}'-{\bm k}_{\perp}''{\bm k}_{\perp}}
\delta_{k_{z}''-k_{z}'k_{z}}
+\delta_{{\bm k}_{\perp}''-{\bm k}_{\perp}'{\bm k}_{\perp}}
\delta_{k_{z}'-k_{z}''k_{z}}
\nonumber \\
&&\hspace{-13mm}
-\delta_{{\bm k}_{\perp}'+{\bm k}_{\perp}''{\bm k}_{\perp}}
\delta_{k_{z}'-k_{z}''k_{z}}
+\delta_{{\bm k}_{\perp}'-{\bm k}_{\perp}''{\bm k}_{\perp}}
\delta_{k_{z}''+k_{z}'k_{z}})\bigr]=0\, ,
\label{BE15c}
\end{eqnarray}
\end{subequations}
with $\tilde{\bm j}_{\perp{\bm k}}$ given by eq.\ (\ref{j_perp-k}).
Finally, entropy in convection is obtained from eqs.\
(\ref{S^2}) and (\ref{T-expand}) as
\begin{eqnarray}
&&\hspace{-10mm}
S^{(2)}_{\rm cv}= -\frac{5}{4}\!\left[
\frac{(\Delta T_{\rm cv})^{2}}{12}-T_{1}^{2}
+\frac{1}{4}\sum_{{\bm k}}
|\tilde{T}_{{\bm k}}|^{2}(1\!+\!\delta_{{\bm k}_{\perp}{\bm 0}})\right.
\nonumber \\
&&\hspace{8mm}\left.
+\Delta T_{\rm cv}\sum_{k_{z}}\tilde{T}_{{\bm k}_{\perp}={\bm 0},k_{z}}
\frac{1\!+\!\cos k_{z}}{k_{z}}\right] .
\label{S^2-cv}
\end{eqnarray}

\section{\label{sec:num}Numerical results}

We now present numerical results on entropy of Rayleigh-B\'enard convection
obtained by solving eqs.\ (\ref{BE15a})-(\ref{BE15c}).
As a model system we consider Ar at $\bar{T}\!=\!273$\,K under atmospheric pressure
and use the values (\ref{parameters-Ar})
for the parameters in eqs.\ (\ref{BE15a})-(\ref{BE15c}).
We also fix the heat flux density $\bar{j}_{Q}$ at $z\!=\!- d/2$
so that temperature difference $\Delta T_{\rm hc}$ 
of the conducting state, expressed 
in terms of $\bar{j}_{Q}$ as eq.\ (\ref{T^1-conducting}),
takes the value $\Delta T_{\rm hc}\!=\! 1$K.
The Rayleigh number $R^{(-1)}$ is then controlled
by changing the thickness $d$. 
We here adopt the units described above eq.\ (\ref{CV}) with $k_{\rm B}\!=\! 1$.

Recent experiments on Rayleigh-B\'enard convection have been performed most actively
with compressed classical gases\cite{Croquette89,deBruyn96,BPA00} 
where controlled optical measurements of convective patterns are possible.
It has been pointed out that the basic $n$ $(\propto \! l^{-1})$ and 
$T$ dependences of eq.\ (\ref{kappa-nu}) are well satisfied
even for those gases under high pressures.\cite{deBruyn96}
Thus, the present consideration with eq.\ (\ref{parameters-Ar})
has direct relevance to those experiments.

\subsection{\label{subsec:NumProc}Numerical procedures}

To solve the coupled equations numerically,
we first multiply eqs.\ (\ref{BE15a})-(\ref{BE15c}) 
by $10^{8}$, $10^{10}$ and $10^{10}$, respectively, 
and rewrite them in terms of 
$\tilde{T}_{\bm k}'\!\equiv\!10^{3}\tilde{T}_{\bm k}$
and $\tilde{j}_{{\bm k}}'\!\equiv\!10^{5}\tilde{j}_{{\bm k}}$
to obtain equations of $O(1)$.
We next replace zeros of 
the right-hand sides by $-\partial\tilde{T}_{\bm k}'/\partial t'$, 
$-\partial\tilde{j}_{z{\bm k}}'/\partial t'$ and
$-\partial\tilde{j}_{p{\bm k}}'/\partial t'$, respectively.
These time derivatives are what one would get by retaining
time dependence in the expansion coefficients of eq.\ (\ref{jT-expand}).
We then discretize the time derivatives as 
$\partial\tilde{T}_{\bm k}'(t')/\partial t'\!
\approx\![\tilde{T}_{\bm k}'(t'\!+\!\Delta t')-
\tilde{T}_{\bm k}'(t')]/\Delta t'$, for example.
The summations over ${\bm k}$
are truncated by using a finite value $\ell_{\rm c}$ $(\gtrsim\! 5)$ in place of $\infty$ 
in eq.\ (\ref{l-range}).
As for periodic structures, we investigate the three 
candidates: the roll, the square lattice
and the hexagonal lattice with $|{\bm b}_{1}|\!= \!|{\bm b}_{2}|\!\sim \!\pi/\sqrt{2}$.
With these preliminaries, we trace time evolution of 
the expansion coefficients until they all acquire constant values.
Choosing $\Delta t'\!\leq\! 0.005$ and $\ell_{\rm c}\!\geq\! 5$
yields excellent convergence for the calculations presented below.
The initial state is chosen as the conducting state 
with small fluctuations $\tilde{T}_{\bm k}'\!\sim\! 10^{-2}$ 
for the basic harmonics ${\bm k}$.
The constants $\Delta T_{\rm cv}$ and $T_{1}$ have been updated at
each time step by using eq.\ (\ref{T_01}).
Also evaluated at each time step is
entropy measured with respect to the heat-conducting state:
\begin{equation}
\Delta S\equiv S^{(2)}_{\rm cv}- S^{(2)}_{\rm hc}\, ,
\label{dS}
\end{equation}
where $S^{(2)}_{\rm hc}$ and $S^{(2)}_{\rm cv}$ are 
given by eqs.\ (\ref{S^2-hc}) and
(\ref{S^2-cv}), respectively.
We thereby trace time evolution of $\Delta S$ simultaneously.
The above procedure is carried out for each fixed periodic structure.

We have studied the range: 
$1\!\leq\! R^{(-1)}/R_{\rm c}\!\leq\! 10$.
Although the region extends well beyond the Busse balloon\cite{Busse78,CH93}
of stability
for classical gases,\cite{Croquette89,deBruyn96,BPA00}
it will be worth clarifying the basic features of steady periodic
solutions over a wide range of the Rayleigh number.

\begin{figure}[b]
\begin{center}
  \includegraphics[width=0.9\linewidth]{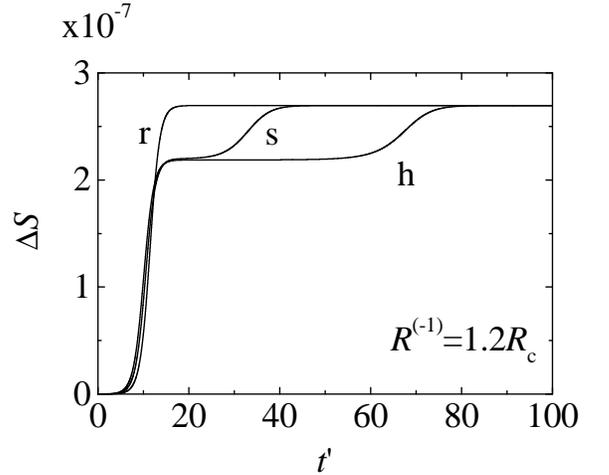}
\end{center}
  \caption{Time evolution of entropy measured with respect to the
  heat-conducting state for $R^{(-1)}\!=\! 1.2R_{\rm c}$. The letters r, s and h denote 
  roll, square and hexagonal, respectively, distinguishing initial fluctuations around
  the heat-conducting solution;
  see text for details. The final state of $t'\!\gtrsim\!80$ is the roll convection,
  whereas the intermediate plateaus of s and h correspond to the square and hexagonal
  convections, respectively.}
  \label{fig:1}
\end{figure}

\subsection{\label{subsec:Num}Results}

Figure \ref{fig:1} plots $\Delta S$ as a function of $t'$
for the Rayleigh number $R^{(-1)}\!=\! 1.2R_{\rm c}$ which is slightly above the
critical value $R_{\rm c}\!=\!27\pi^{4}/4$.
The letters r, s and h denote (r) roll, (s) square and (h) hexagonal,
respectively, distinguishing initial conditions.
Writing $\tilde{T}_{\bm k}'\!=\!\tilde{T}'[\ell_{1},\ell_{2},\ell_{3}]$
and introducing the angle $\theta$ by 
$\theta\!\equiv\cos^{-1}({\bm b}_{1}\cdot{\bf b}_{2})$,
those initial conditions are given explicitly as follows: 
(r) $\tilde{T}'[1,0,1]\!=\!1.00\times 10^{-2}$ with $|{\bm b}_{1}|\!=\!\pi/\sqrt{2}$;
(s) $\tilde{T}'[1,0,1]\!=\!1.01\times 10^{-2}$ and 
$\tilde{T}'[0,1,1]\!=\!0.99\times 10^{-2}$ 
with $|{\bm b}_{1}|\!=\!|{\bm b}_{2}|\!=\!\pi/\sqrt{2}$ and $\theta\!=\!\pi/2$;
(h) $\tilde{T}'[1,0,1]\!=\!\tilde{T}'[0,1,1]\!=\!1.00\times 10^{-2}$ and 
$\tilde{T}'[1,1,1]\!=\!1.01\times 10^{-2}$ 
with $|{\bm b}_{1}|\!=\!
|{\bm b}_{2}|\!=\!\pi/\sqrt{2}$ and $\theta\!=\!2\pi/3$.
We observe clearly that entropy increases
monotonically in all the three cases to reach
a common final value, which is identified as
the roll convection.
Thus, the law of increase of entropy is well satisfied 
even in the open driven system of Rayleigh-B\'enard convection
under the condition of fixed mechanical variables,
i.e., particle number, volume, energy and energy flux.
The intermediate plateaus seen in s and h
correspond to the metastable square and hexagonal lattices,
respectively, with (s) $\tilde{T}'[1,0,1]\!\sim\!\tilde{T}'[0,1,1]$
and (h) $\tilde{T}'[1,0,1]\!\sim\!\tilde{T}'[0,1,1]\!\sim\!\tilde{T}'[1,1,1]$.
The escapes from those lattice structures are
brought about by the tiny initial asymmetry between 
(s) $\tilde{T}'[1,0,1]$ and $\tilde{T}'[0,1,1]$ and (h)
$\tilde{T}'[1,0,1]\!=\!\tilde{T}'[0,1,1]$ and
$\tilde{T}'[1,1,1]$, to form eventually
the roll convection of (s) $\tilde{T}'[0,1,1]\!=\!0$ and
(h) $\tilde{T}'[1,0,1]\!=\!\tilde{T}'[0,1,1]\!=\!0$.
Indeed, those escapes are absent
if we choose (s) $\tilde{T}'[1,0,1]\!=\!\tilde{T}'[0,1,1]$
and (h) $\tilde{T}'[1,0,1]\!=\!\tilde{T}'[0,1,1]\!=\!\tilde{T}'[1,1,1]$
at $t'\!=\!0$.
We have also performed the same calculations for different initial values of
$\tilde{T}'[1,0,1]$, $\tilde{T}'[0,1,1]$, 
$\tilde{T}'[1,1,1]=10^{-6}\!\sim\! 10^{-2}$.
The smaller initial fluctuations merely delay the first steep rise in $\Delta S$
with no observable quantitative changes thereafter.

\begin{figure}[t]
\begin{center}
  \includegraphics[width=0.9\linewidth]{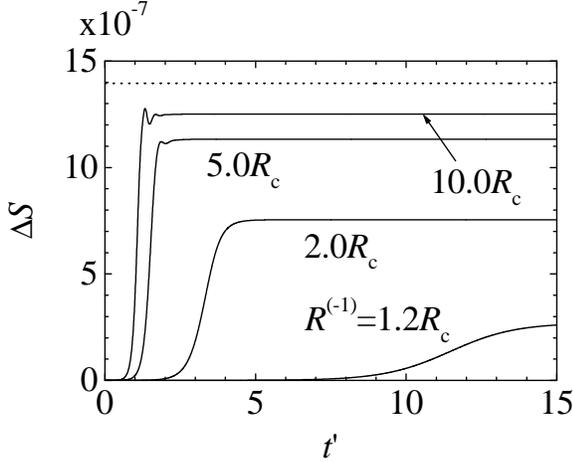}
\end{center}
  \caption{Time evolution of entropy $\Delta S$. 
  The four curves correspond to the different Rayleigh numbers: $R^{(-1)}
  =1.2R_{\rm c}$, $2.0R_{\rm c}$, $5.0R_{\rm c}$ and $10.0R_{\rm c}$.
  The initial state is the
  heat-conducting state with the fluctuation $\tilde{T}'[1,0,1]\!=\!
  1.0\times 10^{-2}$ and ${\bm b}_{1}=\pi/\sqrt{2}$, whereas
  all the final states are the roll convection.
  The broken line near the top indicates the upper bound of $\Delta S$.}
  \label{fig:2}
\end{figure}

Thus, we have confirmed that the roll convection is stable for 
the infinite horizontal area, which was predicted originally
by Schl\"uter, Lortz and Busse\cite{SLB65}
for $R\!\sim \!R_{\rm c}$ based on the linear stability analysis; 
see also refs.\ \onlinecite{Busse78} and \onlinecite{Koschmieder93}.
It should be noted at the same time that the entropy differences 
among different structures are rather small.
We hence expect that: (i) the order of the stability may easily be changed by
finite-size effects, boundary conditions, etc.;
(ii) initial conditions, fluctuations 
and defects play important roles in Rayleigh-B\'enard convection.
These are indeed the features observed experimentally.\cite{Croquette89,deBruyn96,BPA00}

Figure \ref{fig:2} plots time evolution of $\Delta S$ for 
four different Rayleigh numbers, all developing from the 
initial fluctuation
$\tilde{T}'[1,0,1]\!=\!1.00\times 10^{-2}$ with 
$|{\bm b}_{1}|\!=\!\pi/\sqrt{2}$.
Each final state is the roll convection,
which is stabilized faster as $R^{(-1)}$ becomes larger.
The increase in $\Delta S$ is seen quite steep for $R^{(-1)}\!=\! 10R_{\rm c}$
followed by a small oscillation. This oscillation in $\Delta S$ may be due either to
(i) the fluctuations in particle number, momentum,
energy and energy flux inherent to open systems
or (ii) the initial correlations which causes 
the anti-kinetic evolution.\cite{OB67,Cercignani98}
Such fluctuations are also observed in a numerical study
by Orban and Bellemans\cite{OB67,Cercignani98} for an isolated system
and not in contradiction with the law of increase of entropy.
We observe clearly that 
the principle of maximum entropy proposed 
at the beginning of \S\ref{sec:intro}, which is relevant
to the final steady state without time evolution, 
is indeed satisfied here.
The dotted line in Fig.\ \ref{fig:2} is the upper bound of entropy in convection,
as may be realized from eq.\ (\ref{S^2}).
As the Rayleigh number is increased further, entropy differences 
between different structures become smaller so that
the system will eventually fall into the region where fluctuations
dominate with no stable structure.
The turbulence observed in this region\cite{Busse78,CH93,BPA00} 
may be connected with this instability.

\begin{figure}[t]
\begin{center}
  \includegraphics[width=0.9\linewidth]{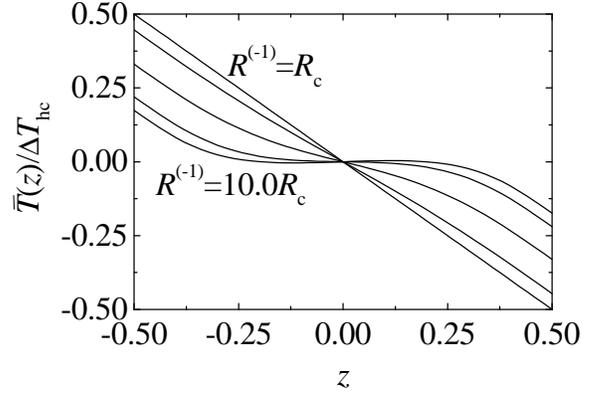}
\end{center}
  \caption{Profile of the average temperature variation $\bar{T}(z)$
  in the roll convection normalized by the temperature difference 
  $\Delta T_{\rm hc}$ in the heat-conducting state. The Rayleigh numbers are
  $R^{(-1)}\!=\! R_{\rm c}$, $1.2R_{\rm c}$, $2.0R_{\rm c}$, $5.0R_{\rm c}$
  and $10.0R_{\rm c}$ from top to bottom on the left part.}
  \label{fig:3}
\end{figure}

Figure \ref{fig:3} shows profile of the average temperature variation
$\bar{T}(z)$ along $z$ in the roll convection for five different
Rayleigh numbers. 
Temperature has less variation as the Rayleigh number becomes larger,
thereby increasing entropy of the system.
Thus, we may attribute the formation of convection to its
efficiency for increasing entropy
under fixed inflow of heat, i.e., the initial slope of temperature.
Experiments on Rayleigh-B\'enard convection have naturally been carried out 
by fixing the temperature difference rather than the inflow of heat, 
and formation of the convection
has been discussed in terms of the change 
in the Nusselt number\cite{Chandrasekhar61,Koschmieder93}
(i.e., the efficiency of the heat transport)
due to the increase of the initial temperature slope through 
the convective transition.
Here we have seen that the same phenomenon can be explained with respect to
the basic thermodynamic quantity of entropy.
Thus, the principle of maximum entropy partly justifies the 
maximum heat transfer hypothesis by Malkus and Veronis.\cite{MV58}

All the above calculations have been carried out by fixing the periodic
structure. 
They clearly show that the principle of maximum entropy is indeed obeyed.
We now take the principle as granted and use it to determine 
the stable lattice structure.
We carry this out within the roll convection
by changing the value $|{\bm b}_{1}|$.

\begin{figure}[t]
\begin{center}
  \includegraphics[width=0.9\linewidth]{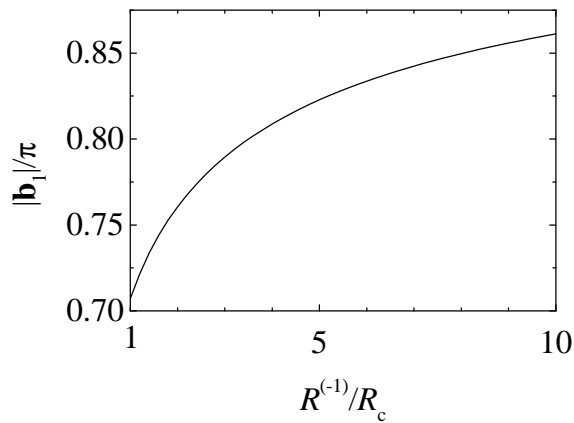}
\end{center}
  \caption{
 The length $|{\bm b}_{1}|$ of the stable roll convection
 as a function of the normalized Rayleigh number $R^{(-1)}/R_{\rm c}$.
 Here $|{\bm b}_{1}|\!=\!\pi/\sqrt{2}$ at $R^{(-1)}/R_{\rm c}\!=\!1$}
  \label{fig:4}
\end{figure}

Figure \ref{fig:4} plots $|{\bm b}_{1}|$ 
as a function of normalized Rayleigh number
$R^{(-1)}/R_{\rm c}$. 
As seen clearly, $|{\bm b}_{1}|$ increases gradually from the value
$\pi/\sqrt{2}$ at $R^{(-1)}/R_{\rm c}\!=\! 1$.
This tendency is in qualitative agreement with the experiment
by Hu, Ecke and Ahlers\cite{HEA93} using a circular cell of
$L/d\!\!\gg \! 1$;
see also ref.\ \onlinecite{BPA00}.
It should be noted at the same time that the entropy difference around 
$|{\bm b}_{1}|\!=\!\pi/\sqrt{2}$ is quite small.
For example, the stable state for $R^{(-1)}/R_{\rm c}\!=\! 10$
corresponds to $|{\bm b}_{1}|\!=\!0.829$, but entropy
is increased by only $2\!\times\! 10^{-9}$ from the value
$\Delta S\!=\!1.296\!\times\! 10^{-6}$ at $|{\bm b}_{1}|\!=\!\pi/\sqrt{2}$.
Hence it is expected that (i) initial conditions, fluctuations and boundary conditions
play important roles and
(ii) it may take a long time, or even be impossible in certain cases,
to reach the stable state.
These are indeed in agreement with experimental 
observations.\cite{Croquette89,deBruyn96,BPA00}

The whole our results have been obtained for the stress-free boundary
condition (\ref{Stress-Free-BC}).
However, the qualitative results will be valid also for more realistic
boundary conditions. 
Indeed, experiments on Rayleigh-B\'enard convection all exhibit
the temperature profile with a steep change near the boundaries
followed by moderate variation in the bulk region
as reflected in the enhancement of the Nusselt number.\cite{Chandrasekhar61,Koschmieder93}
And it is this feature in the present study 
which has caused increase of entropy in convection
over that in the conducting state.

\section{\label{sec:summary}Concluding Remarks}

The present study shows unambiguously that the principle of 
maximum entropy given at the beginning
of \S\ref{sec:intro} is indeed satisfied through 
the Rayleigh-B\'enard convective 
transition of a dilute classical gas.
The result is encouraging for the principle as a general
rule to determine the stability of nonequilibrium steady states.
We need to investigate other open systems,
as well as Rayleigh-B\'enard convection without fixing the lattice structure,
 to confirm its validity further.
 
It may be worth emphasizing once again that entropy/probability,
which is the central concept of equilibrium thermodynamics/statistical mechanics, 
seems to have been
left out of the investigations on nonequilibrium systems and pattern formation.
Indeed, they have been almost always based on deterministic 
equations closely connected with conservation laws.\cite{CH93}
Calculations of entropy for open driven systems imply
treating those finite systems as subjects of statistical mechanics
with considering the boundary conditions explicitly.
Those calculations are expected to shed new light on
nonequilibrium phenomena in general
which are still mysterious.

\begin{acknowledgements}

The author is grateful to H. R. Brand 
for useful and informative discussions
on Rayleigh-B\'enard convection.
This work is supported in part by the 21st century COE program 
``Topological Science and  Technology,'' Hokkaido University.
\end{acknowledgements}

%%%%%%%%%%%%%%%%%%%%%%%%%%%%%%%%%%%%%%%%%%%%%%%%%%%%%%%%%%%%%%%%%%%%%
%%%   The Bibliography
%%%%%%%%%%%%%%%%%%%%%%%%%%%%%%%%%%%%%%%%%%%%%%%%%%%%%%%%%%%%%%%%%%%%%

\end{document}